\UseRawInputEncoding

\documentclass{aastex63}
\usepackage{xcolor}
\usepackage{amsmath}
\usepackage{lipsum}
\usepackage{comment}
\usepackage{makecell}

\usepackage{hyperref}
\usepackage{csvsimple}

\definecolor{darkgreen}{RGB}{0, 100, 0}

\received{December 4, 2020}
\revised{January 6, 2021}
\accepted{January 15, 2021}
\submitjournal{The Astrophysical Journal}

\shorttitle{Massive White Dwarfs in Young Star Clusters}
\shortauthors{Richer et al.}
\graphicspath{{./}{figures/}}

\begin{document}

\title{Massive White Dwarfs in Young Star Clusters}

\correspondingauthor{Harvey B. Richer}
\email{richer@astro.ubc.ca}

\author[0000-0001-9002-8178]{Harvey B. Richer}
\affiliation{University of British Columbia\\
Vancouver, BC, Canada}

\author[0000-0002-4770-5388]{Ilaria Caiazzo}
\affiliation{California Institute of Technology\\
Pasadena, Ca, USA}

\author{Helen Du}
\affiliation{University of British Columbia\\
Vancouver, BC, Canada}

\author{Steffani Grondin}
\affiliation{University of British Columbia\\
Vancouver, BC, Canada}

\author{James Hegarty}
\affiliation{University of British Columbia\\
Vancouver, BC, Canada}

\author[0000-0001-9739-367X]{Jeremy Heyl}
\affiliation{University of British Columbia\\
Vancouver, BC, Canada}

\author{Ronan Kerr}
\affiliation{University of Texas at Austin\\
Austin, Texas, USA\\}

\author{David R. Miller}
\affiliation{University of British Columbia\\
Vancouver, BC, Canada}

\author{Sarah Thiele}
\affiliation{University of British Columbia\\
Vancouver, BC, Canada}

\nocollaboration{9}



\begin{abstract}
We have carried out a search for massive white dwarfs (WDs) in the direction of young open star clusters using the Gaia DR2 database.  The aim of this survey was to provide robust data for new and previously known high-mass WDs regarding cluster membership, to highlight WDs previously included in the Initial Final Mass Relation (IFMR) that are unlikely members of their respective clusters according to Gaia astrometry and to select an unequivocal WD sample that could then be compared with the host clusters' turnoff masses. 

All promising WD candidates in each cluster CMD were followed up with spectroscopy from Gemini in order to determine whether they were indeed WDs and derive their masses, temperatures and ages. In order to be considered cluster members, white dwarfs were required to have proper motions and parallaxes within 2, 3, or 4-$\sigma$ of that of their potential parent cluster based on how contaminated the field was in their region of the sky, have a cooling age that was less than the cluster age and a mass that was broadly consistent with the IFMR. A number of WDs included in current versions of the IFMR turned out to be non-members and a number of apparent members, based on Gaia's astrometric data alone, were rejected as their mass and/or cooling times were incompatible with cluster membership. In this way, we developed a highly selected IFMR sample for high mass WDs that, surprisingly, contained no precursor masses significantly  in excess of ${\sim}$6 $M_{\odot}$.  
 
\end{abstract}



\section{Introduction} \label{sec:intro} 
The maximum mass of a star that is capable of forming a WD is an important astrophysical quantity. It controls the SN II rate, the rate of formation of neutron stars and black holes and the chemical enrichment and star formation rate of galaxies. If the maximum mass is in fact lower than previously thought, the number of supernova explosions would be higher, as would the number of neutron stars and black holes formed. Additionally, the amount of heavy elements pumped back into the interstellar medium would increase along with the dust production and likely the star formation rate in a galaxy. A related quantity is the maximum possible mass of a WD, whose upper limit from theoretical considerations is thought to be 1.38~M$_\odot$ \citep{Chandrasekhar}.
However, presently, there are no WDs known with Gaia parallaxes for which we can say with a high degree of confidence that they have evolved via single star evolution that approach this limit. Single star evolution implies that in any phase of its life the star was not involved in mass transfer from another star or in a merger. The current record-holder for the most massive WD thought to be the product of single-star evolution is GD~50, at 1.28~M$_\odot$. Historically associated with the Pleiades, recent Gaia analysis seems to indicate that GD~50 is not associated with any star cluster but appears to be a member of the 150 Myr old AB Doradus moving group \citep{2018ApJ...861L..13G}. If GD~50 is excluded from the formal IFMR because of its uncertain origin (which makes it difficult to establish its precursor's mass) and the possibility that it may indeed be a merger remnant \citep{1996ApJ...461L.103V}, the most massive single-evolution WD in a star cluster with a well-determined Gaia parallax currently known becomes the Pleiades WD LB1497 at 1.05~M$_\odot$, whose precursor was 5.86~M$_\odot$  \citep{2018ApJ...866...21C}. This WD is well below the Chandrasekhar mass, and therefore the current sample does not come close to testing this upper limit.
It is the goal of our current survey to try and locate more massive single-source WDs.

On a related issue, there appears to be some discrepancy in the rate of supernovae explosions from massive stars. The current upper main sequence mass limit for WD production is thought to be about 8~M$_\odot$ \citep{1983A&A...121...77W}; however, \citet{2011ApJ...738..154H} have shown that, if every star above 8~M$_\odot$ becomes a core-collapse SN II, the predicted rate for type II supernovae would be roughly double the observed rate. One way to resolve this problem would be if the lower mass limit for supernova production were significantly higher: about a 12~M$_\odot$ lower limit to stars exploding as SN II would be needed for a Kroupa IMF \citep{2003ApJ...598.1076K} to bring the prediction in line with observations. We would then expect to see WDs that evolved from main sequence stars as massive as 12~M$_\odot$. At present, no WDs are known with progenitors anywhere near as massive as this, but it is not at all clear whether searches carried out thus far would have been sensitive to these massive WDs, as their high masses would have resulted in small WD radii and hence low luminosities. Additionally, this would have required examining star clusters as young as 20 Myrs for possible WD members, a search that has not been done systematically.

In this survey paper, we remain agnostic about the mass limit for supernova production and we search for WDs in the Gaia DR2 database within open clusters whose turnoff masses are as high as 15~M$_\odot$.

\section{The White Dwarf Survey}  

The Gaia DR2+ catalogue provides a 5-parameter astrometric solution ($\alpha$, $\delta$, $\mu_\alpha$, $\mu_\delta$, $\pi$) for more than 1.3 billion sources. The catalogue is only complete for G magnitudes between $\sim$12 and 17 with a general limiting G magnitude of approximately 21 \citep{gaia_2018}. However, this value can be as bright as G = 18 in dense areas of the sky due to blending from other sources. Considering these restrictions, we primarily surveyed clusters from 10 Myr (turnoff mass $\sim$19.3 M$_\odot$) up to 500 Myr (turnoff mass 2.9 M$_\odot$) within 1~kpc of the Sun. A 1 M$_\odot$ hydrogen atmosphere WD that has been cooling for $\sim$250 Myrs has an absolute G magnitude of $\sim$11.8 \citep{B_dard_2020}, corresponding to an apparent G magnitude of 20.3 at 500~pc. Since young clusters are near the Galactic plane, at this distance we can generally expect about 0.7 mag of extinction, placing the WD close to the Gaia limit of detection.

The current paradigm is that WDs form when a star with mass less than $\sim$8 M$_\odot$ expels its outer layers after the red giant phase, leaving only a core remnant behind. Without fusion occurring, the degenerate core slowly radiates its thermal energy, causing the WD to become increasingly faint and cool. The limitations in the completeness of the Gaia catalogue, coupled with the typical faintness of WDs, results in a bias toward young and nearby open clusters as hosts for massive WDs. Young clusters with higher turnoff masses are more likely to produce higher mass WDs which have not yet cooled beyond Gaia's photometric limit. Additionally, the closer the cluster is to the Sun, the brighter the sources will appear. Nearby clusters can be older and still have observable massive WDs while distant clusters must be very young for potential high-mass WDs to be observable.

\begin{figure}[h]
    \centering
    \includegraphics[width=0.6\textwidth]{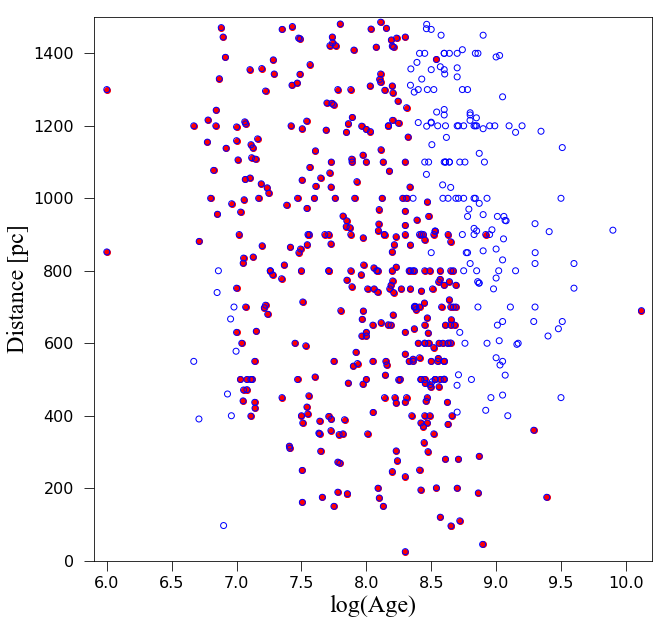}
        \caption{Distance and age of all WEBDA clusters within our survey range (blue circles). The filled circles indicate the clusters that we included in our survey. As can be seen here, the completeness of our survey is very high ($>$ 97\%) for  7.0 $\le$ log (Age) $\le$ 8.5 for distances $\le$ 1500 pc.}
    \label{fig:age-dist}
\end{figure}

The complete list of the 386 surveyed clusters was compiled from WEBDA data. Of this sample, 262 clusters are in the 10-500 Myr age range and are located within 1 kpc of the Sun, 111 have distances of 1-1.5 kpc and ages $<$220 Myr (young + distant) and 8 are in the 500 Myr-2.5 Gyr age range with distances $<$ 400 pc (old + close). The first group is the best suited to look for WDs that might have evolved from massive main sequence stars, but we extended our search to the latter two groups because they also might still contain massive WDs that are observable in Gaia. The distribution of the ages and distances of these clusters are shown in Figure \ref{fig:age-dist}. Also included are the Hyades and NGC 2099 (M37) who do not fit our age restrictions but contain known massive WDs we wished to rediscover. Four extremely young clusters IC 5146, Lynga 14, NGC 6383, Ruprecht 119, with ages $<$10 Myr, were also added as we wanted to remain agnostic to the upper mass limit for WD formation and any WDs present in these clusters could be derived from high-mass progenitors at the upper limit of the IFMR.

We queried the Gaia DR2 archive for each cluster using equatorial cluster centre coordinates as the target and 2$\times$ the angular diameter as the search radius, using data from the DAML02 database \citep{DAML02}. In each cluster, we retrieved the top 500,000 stellar results. Choosing the search radius as twice the literature diameter is an arbitrary choice; we expect that it should allow for the majority of cluster members to be included, accounting for a possible underestimate in the literature diameter while not searching too widely and flooding the data with non-cluster stars. No further preliminary cuts were made to reduce contamination.

\section{Determining Cluster Parameters}

In tandem with the search for massive WDs, new cluster parameters were determined for every identifiable cluster to allow for membership determination and follow-up analysis. We identified cluster members through cuts in positional space, parallax and proper motions, assuming that the probability distributions are well approximated by Gaussian probability density functions (PDFs): 2-dimensional in proper motion and equatorial positional space and 1-dimensional in parallax. Reported cluster centres and errors in our survey table (Table \ref{tableclusters}) correspond to the means and standard deviations of the Gaussians. 

\begin{figure}[h]
    \centering
    \begin{tabular}{cc}
    \includegraphics[width=65mm]{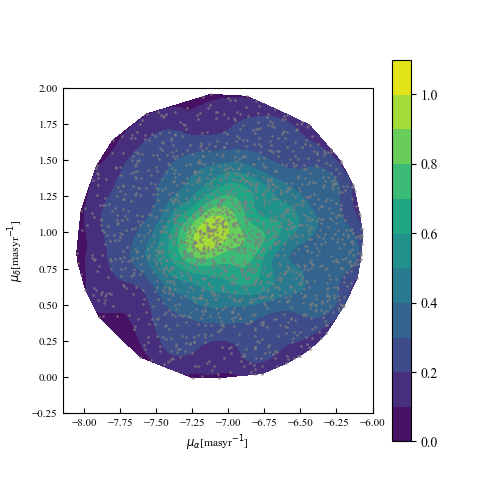} &  \includegraphics[width=65mm]{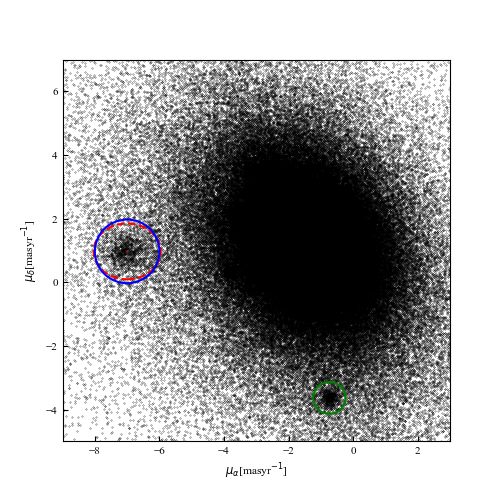} \\
    \includegraphics[width=65mm]{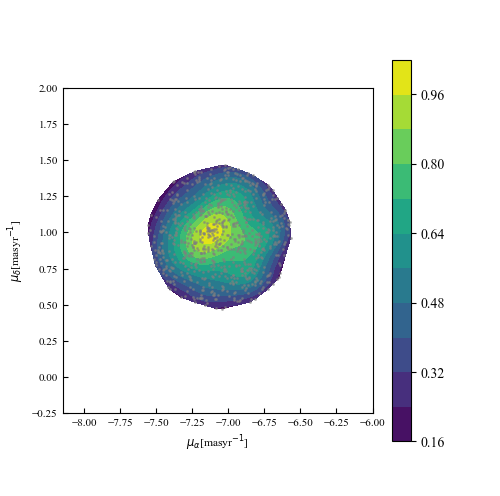}  & \includegraphics[width=65mm]{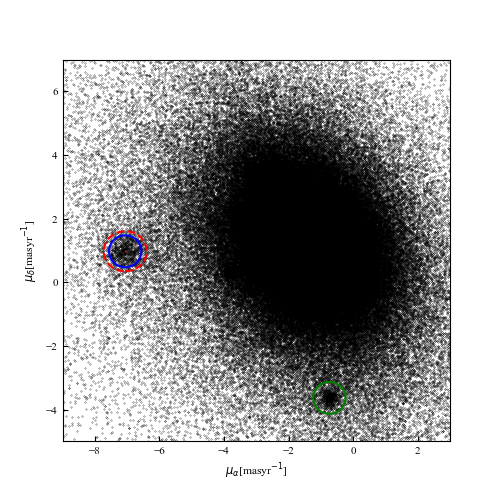}\\
    \end{tabular}
\caption{KDE colormap with $h$=0.1 (left) and proper motion selection (right) for M47 (NGC 2422). The KDE was generated from a cut in proper motion with a radius 1 mas yr$^{-1}$ (top) and 0.5 mas yr$^{-1}$ (bottom), also indicated by the solid blue line in the right-hand plots. A 2$\sigma$ ellipse from the resulting fit is shown as a red dashed line. Also visible is the large, dense region of contaminating stars from the galactic disk and NGC 2423 (below, circled in green). From this figure it is clear that clusters are required to have a significant peculiar motion with respect to the galactic disk to be separated by a proper motion cut and avoid important contamination.}
\label{fig:KDE}
\end{figure}

We fitted Gaussian functions to kernel density estimates (KDEs) in each space. KDEs are used as a method to obtain a density map for a collection of coordinates\footnote{We use the scikit-learn’s KernelDensity class in Python \citep{scikit-learn}.}. We chose to use KDEs instead of histograms for the purpose of fitting Gaussian models to density distributions for three main reasons. Firstly, discretely binned data leads to a loss of information, whereas in a KDE, each coordinate is uniquely assigned a density. This provides the model fitting algorithm with a larger number of points to improve the quality of the fit, which is even more important when working with open star clusters that can be very sparse, with sometimes less than 100 members. 
Secondly, the number of bins in an histogram needs to be properly determined for each data set, 
while KDEs are determined by two parameters: kernel and bandwidth $h$. We chose a Gaussian kernel, which is the most common choice, for which the bandwidth $h$ is equivalent to the standard deviation $\sigma$. We found the `best' bandwidth to be always close to 0.1 in proper motion and parallax and close to 0.5 in position\footnote{The `best' bandwidth was determined by scikit-learn’s GridSearchCV algorithm in Python \citep{scikit-learn}.}.
Finally, the KDE is useful as a visual aid. We show an example in Figure~\ref{fig:KDE}: when the radius of the cut in proper motion is too large (upper panels), we see an excess density of stars towards the right hand side of the KDE because of contamination from the galactic disk, which causes the fitted standard deviation to be overestimated. The distortion in the KDE in the direction of the contamination from the field is helpful to iteratively adjust the subset of candidate stars chosen for fitting, to find a balance between including as many cluster members as possible and reducing the contamination.
\textbf{}

The limitation of this manually conducted survey is that a cluster needs to be clearly identified in proper motion space (see Fig.~\ref{fig:KDE}). Of the total sample of 386 clusters, 219 were distinguishable in proper motion space and 167 were not. \cite{cantat-gaudin-2018} used Gaia data to derive cluster parameters for 1229 nearby open clusters using the unsupervised membership code \texttt{UPMASK}. Whenever the identification of clusters in contaminated and crowded fields was not possible visually in proper motion space, we used this catalog instead. However, out of 169 clusters which were not identifiable by eye in proper motion, 110 also do not appear in \cite{cantat-gaudin-2018}. 


We applied astrometric cuts in proper motion and parallax space, assuming the cluster population to be well described in each dimension by a Gaussian PDF. 
We find that retaining only stars that lie within 2-$\sigma$ of the mean proper motion and parallax of the cluster strikes the best balance between retaining as many `real' cluster member as possible and avoiding contamination from field stars. Employing a 3-$\sigma$ cut instead leads to an important contamination, which affects later analysis such as subsequent fit quality, cluster reddening, extinction and cluster age determination. Therefore, only sources within 2-$\sigma$ of the mean proper motion and parallax are considered likely cluster members. However, due to typically high uncertainty in Gaia proper motion and parallax data for faint sources, once cluster parameters were fully determined, we tested 4-$\sigma$ cuts to probe the presence of WDs farther from the cluster centres whose error bars might intersect the 2-$\sigma$ region even if nominal values do not.

After making 2-$\sigma$ cuts in proper motion space, we analyzed KDEs in parallax space. We assumed that the parallax data could be described by the sum of two 1-D Gaussians: one describing the distribution of the field sources (noise) and one the distribution of the cluster. Fits for these two Gaussians were performed on the KDE. The parallax centre for the noise was typically in the range 0.1-0.5 mas. If the cluster is `nearby' and has a parallax centre $\ge$1.5 mas, the Gaussian describing the noise could be omitted from the fitting. Describing the data with a pair of Gaussians becomes extremely useful for clusters whose parallax centre is $\le$ 1.5 mas as the noise does not displace the fitted parallax centre and does not cause the fitted $\sigma$ to be overestimated. We then retain all the stars that are within 2 $\sigma$ from the mean. After making cuts in parallax space, we have completed all astrometric cuts. We then fit a Gaussian to a KDE of the equatorial position distribution of remaining sources.

We carry out photometric filtering by removing all objects with the photometric excess flag (\texttt{phot\_bp\_min\_rp\_excess\_factor}), $E$ $\geq$ 1.5. $E$ is the sum of $G_{BP}$ (\texttt{phot\_bp\_mean\_flux}) and $G_{RP}$ (\texttt{phot\_rp\_mean\_flux}) over the total $G$ band flux (\texttt{phot\_g\_mean\_flux}) and is expected to be larger than 1 by a small factor, with wide deviation suggesting lower photometric precision \citep{Evans_2018}. \cite{lindegren-2018} has recommended keeping sources which satisfy
\begin{equation}
    1 + 0.015(G_{BP}-G_{RP})^2 < E <1.3 + 0.065(G_{BP}-G_{RP})^2.
	\label{excess}
\end{equation}
 
The upper bound of equation (1) removes very faint sources that might have been rejected by a filtering on the photometric precision \citep{Arenou_2018}. These sources appear most abundantly in the galactic plane, where many of our surveyed clusters lie. Furthermore, faint sources with imprecise photometry could be mistaken for interesting WDs. The value of 1.5 was selected as the upper threshold for all clusters rather than filtering sources individually. A lower cut-off as given in equation (1) is primarily needed to filter perturbations by close-by sources \citep{Arenou_2018} for observing very distant sources, but typically only removes a small number and was deemed unnecessary for our survey.

To determine the cluster reddening, we considered only cluster members that had complete reddening and extinction data from the Gaia DR2 catalogue. Due to a lack of adequate parallax and photometric precision, Gaia DR2 does not determine astrophysical parameters for sources with $G \geq 17.068766$ \citep{Andrae_2018}. For the remaining sources \cite{Andrae_2018} use distances determined via parallax and $G$, $G_{\rm BP}$, and $G_{\rm RP}$ magnitudes as inputs to train the machine learning algorithm Extra-Trees \citep{Geurts_2006}, which, when combined with extinction models, allows it to compute the $G$-band extinction, $A_{\rm G}$, and the reddening, $E$(BP-RP) $= A_{\rm BP} - A_{\rm RP}$, for each source. Extra-Trees is unable to extrapolate beyond the intervals of the training variable, as a result they avoid negative estimates for $A_{\rm G}$ and $E$(BP-RP). This leads to some sources being given zero as their 1$\sigma$ lower-bound extinctions. For consistency, we removed these sources from our reddening calculations. Additionally, a number of sources suffered from degeneracies due to model stars with varying astrophysical parameters having identical apparent magnitudes.  Extra-Trees cannot distinguish between these, resulting in some sources having very large uncertainty intervals. Sources whose extinction and reddening uncertainty intervals failed to meet certain conditions were not retained in Gaia DR2 \citep{Andrae_2018}.  

Based on these constraints, only $\approx 42\%$ of the sources in our samples after the astrometric cuts had complete reddening and extinction data. Using these sources we calculated the variance-weighted means to estimate the overall reddening and extinction of each cluster. The error in these estimates was found by first determining the reliability weight-based unbiased sample variance, which was used alongside the number of sources to determine the standard error in each mean. Though this helps to account for the increased statistical error in the clusters with relatively few members, we emphasize that some of these values were calculated based on just a handful of sources and may not be overly reliable. 

With cluster reddening and extinction determined, an extinction, colour and distance corrected colour-magnitude diagram could then be constructed to identify any potential WDs, their cooling times, and the cluster age.

We note that in the late stages of preparation of this manuscript, a related survey appeared on the archives \citep{2020arXiv201103578P} that carried out a similarly motivated search for cluster WDs. Aside from some differences in methodology and scope, one important difference between our search and that one is that we have verified many of our best candidates with spectra, a critical component in establishing the properties of candidate cluster WDs. 

\section{Expected Number of White Dwarfs in Selected Star Clusters} 
\label{sec:expectedNwd}
Although WDs are the final evolutionary point for ninety-eight percent of all the stars in our galaxy, studies have shown that the observed number of WDs in many clusters is significantly smaller than expected \citep{kalirai_2009, 2003ApJ...595L..53F}. A possible explanation of this discrepancy are natal WD velocity kicks, occurring if the red giant precursor asymmetrically ejects its outer layers, driving the WD in the opposing direction.  These kicks could potentially eject a WD from its parent cluster, resulting in a deficiency of observed WDs \citep{2003ApJ...595L..53F}. 

While a total of 387 clusters were surveyed for massive WDs in this work, here we present an estimate of the number of WDs {\it expected} in each cluster, $\langle{N_{WD}}\rangle$, for 163 of the clusters in our sample with a well-defined turnoff (so that isochrones could be reliably fit) and good reddening values (generally $\sigma$E(BP-RP)$\le 0.05$). Using the same filtered data files as in our massive WD search, we first produced CMDs for each cluster in order to determine its age. We converted the Gaia apparent magnitudes to absolute magnitudes using distances determined by cluster centre parallaxes and extinctions derived from reddening values ($E(BP-RP)$). While the CMDs act as an excellent visual aid when determining cluster ages, it is the information contained within the isochrone of the associated age that is key in determining  $\langle{N_{WD}}\rangle$. All isochrones in this study were obtained from the STEV CMD server\footnote{http://stev.oapd.inaf.it/} and use Gaia DR2's photometric system and passbands from \cite{2018A&A...619A.180M}. Solar metallicity was used for all isochrones.

To expedite the age determination, when available, we used the ages found by machine learning in \citet{2020arXiv200407274C} as our first guess for isochrone fitting. We then visually inspected how well the isochrones reproduced the data on the Gaia CMD. Whenever the fitting was satisfactory, we kept the same ages as in \citet{2020arXiv200407274C}; as the authors mention that the error in $\log(t)$ is between 0.15-0.25 for their method, we employ an error of 0.20 in $\log(t)$ for each cluster, where $t$ is the age in Myr. For the clusters that are absent from \citet{2020arXiv200407274C} or for which the first-guess isochrone fits the cluster data poorly, we manually fit isochrones on each cluster to find the age. Cluster-specific factors such as contamination by field stars and poor definition of the turnoff point affect the uncertainty in our visual age determinations. These uncertainties were computed by manually adjusting the age of the isochrone to best fit the turnoff stars of the cluster, finding a maximum and a minimum age for which the isochrone is still a good fit. The value we quote for the age is then the midpoint between the two, and the error is half the difference. It is also important to note that we were unable to determine ages and thus WD number expectation values for some of the clusters. Reasons for this include poorly determined observational parameters, highly contaminated clusters and the absence of sufficient turnoff stars. Cluster ages and uncertainties are indicated in Table \ref{tableclusters}, where a $\dagger$ denotes an age from \cite{2020arXiv200407274C}, $\ddagger$ indicates manual age determination from this work and * indicates that no age determination was possible. 

For each cluster, from the isochrone, we assume a \citet{2001MNRAS.322..231K,2002Sci...295...82K} initial mass function (IMF), \emph{i.e.} the initial distribution of masses of the stars in the cluster $\mathrm{d}N/\mathrm{d}M$. From the IMF we could compute the number of WDs that have already been produced by the cluster, or $\langle{N_{WD}}\rangle$, by integrating the IMF from the mass of the stars that are currently producing white dwarfs ($M_{\rm init, WD}$) to a chosen upper mass limit for the production of white dwarfs (in particular, we used 8~M$_\odot$). Of course we had to normalize this number to the actual number of stars in the cluster. In order to do so, we used the brightest main-sequence stars: we sorted all cluster stars by brightness and obtained the number of main-sequence stars in the brightest third of the cluster ($N_{1}$), up to the turn-off, and divided it by the number of stars in the same range predicted by the IMF, \emph{i.e.} we divided $N_{1}$ by the integral of $\mathrm{d}N/\mathrm{d}M$ between the mass of the faintest of the selected stars ($M_{\rm init, 3}$) and the mass of the turn-off ($M_{\rm init, TO}$). The turnoff point for each cluster was identified by finding the highest temperature value reached by stars on the main sequence. We chose to focus on the brightest one-third of stars as a trade-off between having more stars and hence better statistics and the fact that lower mass stars may have already been expelled from the cluster through dynamical interactions. The number of expected WDs for each cluster was therefore obtained as:

\begin{equation}
    \left < {N_{WD}} \right >=N_1 \left [ \int_{M_{\rm init,3 }}^{M_{\rm init, TO}} \frac{\mathrm{d}N}{\mathrm{d}M} \;\mathrm{d}M \right ]^{-1} \int_{M_{\rm init, WD}}^{8M_\odot} \frac{\mathrm{d}N}{\mathrm{d}M}\;\mathrm{d}M.
	\label{eq:expectednumber}
    \end{equation}

Regardless of whether the age was determined here or obtained from \cite{2020arXiv200407274C}, the error in $\langle{N_{WD}}\rangle$ was computed from the error on the age, by taking the absolute value of the difference between the number of expected white dwarfs for the nominal age plus the error and minus the error and dividing by two. Since these errors arise directly from the uncertainty in the cluster age, the uncertainty in $\langle{N_{WD}}\rangle$ varies dramatically from cluster to cluster. The $\langle{N_{WD}}\rangle$ and corresponding uncertainties are presented in Table \ref{tableclusters}. The total number of white dwarfs that we expect to have been produced by these clusters is nearly 1,100 if the maximum initial stellar mass for white dwarf formation is $8M_\odot$.  On the other hand, if white dwarfs are produced from stars only up to $6M_\odot$, the expected number of white dwarfs is 30\% smaller at 750.  Depending on the age of the cluster, the effect of the upper mass for white-dwarf formation can be modest (about 10\%) or large (a factor of several).  However, because the expected number of white dwarfs is dominated by the richer clusters which happen to be older as well, the net effect is a 30\% decrement.  Increasing the upper limit on white dwarf formation to an initial stellar mass of $12M_\odot$ increases the number of expected white dwarfs by 30\% to nearly 1,400.



\section{Survey Results: Astrometry, Photometry and Spectroscopy of WD Candidates}

In our narrow search (a 2$\sigma$ cut in proper motion and parallax), we identified twenty-four WD candidates in the direction of young open star clusters. Of these twenty-four, thirteen were already known cluster members, well-studied in the literature (see \cite{2018ApJ...866...21C,  2016ApJ...818...84C,  2013AJ....145..134C, 2012MNRAS.423.2815D, 2004MNRAS.355L..39D, 2011ApJ...743..138G, 2008ApJ...676..594K}) and another five have masses, inferred from their location in the Gaia CMD, that appear inconsistent with cluster membership, leaving six new potential high mass WD candidates. Spectra of each of these six candidates were obtained with Gemini North or South in order to determine their surface gravity, temperature and hence their cooling age and eventually their mass. All this data together allowed for a critical assessment of the cluster membership of each candidate.  As mentioned above, because of the high uncertainty in Gaia proper motion and parallax data for faint sources, we relaxed our parameters to search for white dwarfs that are within 4-$\sigma$ from the cluster centres in either parallax or proper motion and we discovered an additional four high mass WD candidates, that we followed-up with spectra from Gemini. 

A handful of well-studied WDs met our proper motion and parallax criteria but did not appear in our final narrow search results. Two of these were missed because their excess factors were too high, while another five did not appear in the original Gaia queries. The WDs which missed the queries likely missed as a result of being in a particularly nearby cluster with a wide angular extent, in which case the 500,000 returned results was insufficient. The outstanding example of this was the Hyades Cluster, which covers at least 100 square degrees on the sky. It would not be possible to query the Gaia database to include the enormous number of stars in this area. It is possible this issue may have led to us missing additional WDs in other near and large clusters, but we expect the impact to be minor at worst. Table~\ref{tableclusters-all} lists all of the identified candidate WDs along with the handful of well-studied WDs appearing in \cite{2018ApJ...866...21C} which were missed for the aforementioned reasons. The $\dagger$ in the comments indicates that the WD passed the 4-$\sigma$ but not 2-$\sigma$ cuts, $\ddagger$ is for WDs which passed 2-$\sigma$ cuts but had excess factors which were too high, and the $*$ signifies WDs which were missed by the Gaia queries entirely. The remaining objects listed in \cite{2018ApJ...866...21C} that did not make our cuts are included in Table~\ref{tableWDs-excluded} in the appendix.

This section provides the details of the astrometric search in the direction of each of the ten clusters that we discovered to potentially host a new massive WD. As part of the process, we develop the cluster's CMD, fit isochrones \citep{2012MNRAS.427..127B}
to it in order to estimate the cluster age, and include WD cooling sequences in the CMD in order to photometrically estimate the mass and cooling age of the WD. We also analyse the Gemini spectra of each of the WDs and fit spectral models to the Balmer absorption lines, deriving the WD's surface gravity ($\log g$) and effective temperature ($T_{\rm{eff}}$).

Our fitting method to the Balmer lines is similar to the routine outlined in \citet{2005ApJS..156...47L}: we fit the spectrum with a grid of spectroscopic models combined with a polynomial in $\lambda$ (up to $\lambda^9$) to account for calibration errors in the continuum; we then normalize the spectrum using this smooth function picking normal points at a fixed distance in wavelength to the lines and finally use our grid of model spectra to fit the Balmer lines and extract the values of $\log g$ and $T_{\rm{eff}}$. The nonlinear
least-squares minimization method of Levenberg-Marquardt is used in all our fits. We employ the models of pure hydrogen atmospheres developed by \citet{2011A&A...531L..19T} for most of the objects and the hydrogen atmospheres polluted by metals developed by \citet{2010ApJ...720..581G} for the hot WD in ASCC 47. This WD is so hot ($T_{\rm{eff}}$ $>$ 110,000~K) that the levitation of metals caused by the resultant high radiation pressure cannot be ignored. These fits are shown in Fig.~\ref{fig:lines} and the resulting parameters are listed in Table~\ref{tab:spec}. 

A complete listing of all the clusters surveyed for massive WDs can be found in Table \ref{tableclusters}. The ten clusters, together with their ages as derived from our new analysis of Gaia DR2 data, are listed in Table~\ref{tableclusters2}. For all the candidates we provide the WD mass as well as that of its precursor if the WD appears to be a member of its respective cluster. As we shall see, several WD candidates turned out to be probable foreground or background objects and a few were not even WDs at all!

\begin{deluxetable*}{lclccccc}
\tablecaption{Potential Cluster Member White Dwarfs Identified\label{tableclusters-all}}
\tablewidth{700pt}
\tabletypesize{\scriptsize}
\tablehead{
\\
\multicolumn{2}{c}{Cluster} 
& \multicolumn{6}{c}{WD} 
\\
\multicolumn{2}{c}{\rule{4cm}{.02cm}}
& \multicolumn{6}{c}{\rule{13cm}{.02cm}}
\\
\colhead{Name} &
\colhead{$E\rm{(Bp-Rp)}$} &
\colhead{Gaia Source ID} &
\colhead{G$_{obs}$} &
\colhead{(Bp - Rp)$_{obs}$} &
\colhead{G$_{0}$} &
\colhead{(Bp - Rp)$_0$} &
\colhead{Comments}
}
\startdata
Alessi 8     & 0.149$\pm$0.022 & 5888965556642170624$^+$ & 20.818$\pm$0.013 & -0.334$\pm$0.140 & 11.360$\pm$0.228 & -0.483$\pm$0.141 & followed-up \\
Alessi 21    & 0.100$\pm$0.011 & 3050806942132626048$^+$ & 20.360$\pm$0.001 &	-0.293$\pm$0.140 & 11.357$\pm$0.018 & -0.393$\pm$0.140 & followed-up$\dagger$ \\
ASCC 47      & 0.136$\pm$0.015 & 5529347562661865088     & 18.714$\pm$0.003 & -0.509$\pm$0.042 &  8.957$\pm$0.241 & -0.645$\pm$0.045 & followed-up \\
ASCC 113     & 0.078$\pm$0.006 & 1871306874227157376     & 19.833$\pm$0.004 & -0.435$\pm$0.107 & 10.889$\pm$0.010 & -0.513$\pm$0.107 & followed-up$\dagger$ \\
M39          & 0.044$\pm$0.008 & 2170776080281869056     & 19.193$\pm$0.003 &	-0.179$\pm$0.050 & 11.732$\pm$0.015 & -0.222$\pm$0.051 & followed-up$\dagger$ \\
M44          & 0.096$\pm$0.013 & 5597682038533250304     & 19.567$\pm$0.004 & -0.075$\pm$0.073 & 10.309$\pm$0.172 & -0.171$\pm$0.074 & low mass [1]    \\
M44          & 0.119$\pm$0.009 & 659494049367276544      & 18.215$\pm$0.002 & -0.134$\pm$0.024 & 11.603$\pm$0.367 & -0.253$\pm$0.024 & well-studied [3]$*$ \\
M44          & 0.119$\pm$0.009 & 661010005319096192      & 18.356$\pm$0.001 & -0.075$\pm$0.017 & 11.743$\pm$0.367 & -0.194$\pm$0.017 & well-studied [3]$*$ \\
M44          & 0.119$\pm$0.009 & 661270898815358720      & 17.682$\pm$0.002 & -0.290$\pm$0.025 & 11.069$\pm$0.160 & -0.409$\pm$0.027 & well-studied [3] \\
M44          & 0.119$\pm$0.009 & 661297901272035456      & 18.413$\pm$0.002 & -0.093$\pm$0.025 & 11.114$\pm$0.160 & -0.212$\pm$0.021 & well-studied [3] \\
M44          & 0.119$\pm$0.009 & 661311267210542080      & 17.682$\pm$0.002 & -0.024$\pm$0.025 & 11.780$\pm$0.160 & -0.143$\pm$0.019 & well-studied [3] \\
M44          & 0.119$\pm$0.009 & 661353224747229184      & 17.883$\pm$0.001 & -0.168$\pm$0.017 & 11.271$\pm$0.160 & -0.287$\pm$0.019 & well-studied [3] \\
M44          & 0.119$\pm$0.009 & 661841163095377024      & 18.342$\pm$0.001 & -0.056$\pm$0.018 & 11.729$\pm$0.367	& -0.175$\pm$0.018 & well-studied [3]$*$ \\
M44          & 0.119$\pm$0.009 & 662798086105290112      & 17.960$\pm$0.002 & -0.171$\pm$0.017 & 11.348$\pm$0.367 & -0.290$\pm$0.017 & well-studied [3]$*$ \\
M44          & 0.119$\pm$0.009 & 664325543977630464      & 17.999$\pm$0.002 & -0.141$\pm$0.033 & 11.387$\pm$0.160 & -0.260$\pm$0.035 & well-studied [3] \\
M44          & 0.119$\pm$0.009 & 665139697978259200      & 18.026$\pm$0.002 & -0.121$\pm$0.033 & 11.413$\pm$0.367 & -0.240$\pm$0.033 & well-studied [3]$*$ \\
M47          & 0.149$\pm$0.006 & 3029912407273360512     & 19.796$\pm$0.005 & -0.134$\pm$0.106 & 11.071$\pm$0.233 & -0.283$\pm$0.106 & followed-up \\
NGC 2516     & 0.089$\pm$0.004 & 5290834387897642624     & 19.165$\pm$0.002 & -0.261$\pm$0.069 & 10.889$\pm$0.297 & -0.350$\pm$0.069 & well-studied [2] \\
NGC 3114     & 0.134$\pm$0.004 & 5256341647300346496$^+$ & 20.929$\pm$0.018 & -0.450$\pm$0.532 & 10.537$\pm$0.290 & -0.584$\pm$0.532 & followed-up \\
NGC 3532     & 0.061$\pm$0.006 & 5340154429370971648$^+$ & 20.703$\pm$0.012 &  0.353$\pm$0.394 & 12.046$\pm$0.497 &  0.292$\pm$0.394 & low mass [1]     \\
NGC 3532     & 0.061$\pm$0.006 & 5338636244376571136$^+$ & 20.093$\pm$0.006 &  0.148$\pm$0.165 & 11.437$\pm$1.144 &  0.087$\pm$0.165 & well-studied [2]$\ddagger$ \\
NGC 3532     & 0.061$\pm$0.006 & 5340149103605412992$^+$ & 20.189$\pm$0.007 & -0.382$\pm$0.235 & 11.533$\pm$1.144 & -0.443$\pm$0.255 & well-studied [2]$\ddagger$ \\
NGC 6087     & 0.292$\pm$0.012 & 5832123141956843648$^+$ & 20.753$\pm$0.007 &  0.123$\pm$0.360 & 10.189$\pm$0.530 & -0.169$\pm$0.360 & low mass [1]   \\
NGC 6633     & 0.244$\pm$0.018 & 4477214475044842368     & 18.914$\pm$0.004 &  0.014$\pm$0.091 & 10.426$\pm$0.214 & -0.230$\pm$0.093 & low mass [1]    \\
Pleiades     & 0.096$\pm$0.005 & 66697547870378368       & 16.614$\pm$0.001 & -0.431$\pm$0.013 & 10.745$\pm$0.149 & -0.527$\pm$0.013 & well-studied [4] \\
Ruprecht 147 & 0.202$\pm$0.016 & 4087806832745520128     & 18.836$\pm$0.004 & -0.167$\pm$0.069 & 10.974$\pm$0.130 & -0.369$\pm$0.070 & well-studied [5] \\
Ruprecht 147 & 0.202$\pm$0.016 & 4183919237232621056     & 18.688$\pm$0.004 &  0.004$\pm$0.057 & 10.826$\pm$0.130 & -0.198$\pm$0.060 & well-studied [5] \\
Ruprecht 147 & 0.202$\pm$0.016 & 4183928888026931328     & 18.803$\pm$0.003 &  0.012$\pm$0.066 & 10.941$\pm$0.130 & -0.191$\pm$0.068 & well-studied [5]\\
Ruprecht 147 & 0.202$\pm$0.016 & 4183937688413579648$^+$ & 19.106$\pm$0.004 &  0.073$\pm$0.128 & 11.243$\pm$0.130 & -0.130$\pm$0.129 & well-studied [5]\\
Ruprecht 147 & 0.202$\pm$0.016 & 4184148073089506304     & 19.628$\pm$0.005 &  0.028$\pm$0.084 & 11.765$\pm$0.130 & -0.174$\pm$0.085 & well-studied [5]\\
Ruprecht 147 & 0.202$\pm$0.016 & 4184169822810795648     & 18.894$\pm$0.005 & -0.021$\pm$0.087 & 11.032$\pm$0.130 & -0.223$\pm$0.088 & well-studied [5]\\
Stock 2      & 0.378$\pm$0.009 & 506862078583709056      & 19.646$\pm$0.005 & -0.080$\pm$0.088 & 10.978$\pm$0.133 & -0.458$\pm$0.089 & followed-up \\
Stock 2      & 0.378$\pm$0.009 & 507362012775415552      & 19.794$\pm$0.005 &  0.175$\pm$0.105 & 11.126$\pm$0.133 & -0.203$\pm$0.106 & low mass [1]\\
Stock 12     & 0.140$\pm$0.012 & 1992469104239732096     & 19.099$\pm$0.003 & -0.126$\pm$0.055 & 10.581$\pm$0.019 & -0.266$\pm$0.056 & followed-up$\dagger$\\
vdB Hagen 23 & 0.108$\pm$0.026 & 5541515415474844544     & 20.269$\pm$0.006 & -0.308$\pm$0.170 & 11.776$\pm$0.185 & -0.416$\pm$0.172 & followed-up \\
\enddata
{\medskip {\bf Notes.} $^+$ indicates the object does not appear in the \cite{Gentile_Fusillo_2018} catalog. $\dagger$ indicates the object fell just outside the narrow search range, i.e. a 2-$\sigma$ cut in parallax and proper motion of the cluster but was nevertheless included in our sample of followed-up WDs. $\ddagger$ indicates an object that passed the 2-$\sigma$ cuts but had an excess factor $>1.5$, while $*$ indicates the object missed the original Gaia queries, these well-studied objects were included for completeness. [1] WD appeared to be below 0.6~$M_{\odot}$ in the cluster CMD and was judged to be a non-member and was not pursued further. [2] \cite{2012MNRAS.423.2815D}, [3] \cite{2004MNRAS.355L..39D}, [4] \cite{2011ApJ...743..138G}, [5] \cite{2013AJ....145..134C} but initial mass below 2.5~$M_{\odot}$ and cluster too old (\cite{2020NatAs...4.1102M}).}
\end{deluxetable*}

\subsection{vdB Hagen 23}
\begin{figure}[tb]
    \centering
    \includegraphics[width=0.99\textwidth]{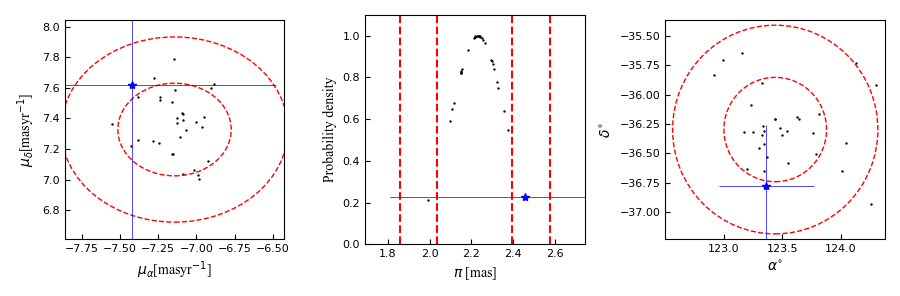}
    \caption{Astrometric data for vdB-Hagen 23. Proper motion (left), KDE of parallax values (centre) and equatorial position (right). Red dashed lines indicate 1 and 2 sigma contours. Blue star with error bars indicates the WD candidate values.}
    \label{fig:vdbh23}
\end{figure}

vdB Hagen 23 is an extremely sparse cluster with a WEBDA age of only 14 Myrs. From the CMD Fig.~\ref{fig:3panel1} (leftmost panel), it is clear that the cluster is very young (our best estimate is 55$\pm20$ Myr) and the WD, if really a cluster member, appears massive ($\sim$1.1~M$_\odot$). It is surprising and remarkable that such a young cluster seems to potentially host a WD member. If confirmed this would mean a precursor mass of about 15~M$_\odot$. As can be seen in Fig.~\ref{fig:vdbh23}, the WD candidate is well within the 2$\sigma$ limits of all the astrometric criteria, though with rather large errors in all these three quantities.

\textbf{}

From the position of the WD in the CMD, however, the WD appears much too faint and cool: cooling models from the Montreal group 
(\citet{2010ApJ...720..581G})
indicate that the time for a WD of about 1.1~M$_\odot$ to cool to the absolute G magnitude that we obtain using the parallax of the WD (11.775) and the observed color ($Bp - Rp = -0.416$) is about 190 million years, significantly older than the  cluster. 
The analysis of the spectrum of the WD (Figs.~\ref{fig:totspec} and~\ref{fig:lines}), provides a surface gravity and an effective temperature (Table~\ref{tab:spec}) that imply a WD mass near 0.63~M$_\odot$, in strong disagreement with the photometric estimate. If we apply the cluster parallax to the WD instead of its own parallax, this yields a WD that is only 0.2 magnitudes brighter, not nearly enough to explain the difference. The only reasonable way in which to bring these contradictory estimates into agreement is to assert that the WD is in fact much closer to us than the cluster, and hence unrelated to it. The WD would need to have a parallax that is even larger than its 2$\sigma$ limit to explain its absolute magnitude --- a parallax that puts it at about half the cluster distance.
 
We conclude from this discussion that the WD in the direction of vdB-Hagen 23 is not a cluster member, even though the astrometric data make a plausible case for membership.
 This example illustrates that, even with our very strict requirements for cluster membership, the large errors in parallax and proper motion of the WDs, due to the fact that they are rather faint objects, often translate into a reasonable possibility of finding false-candidates. For this reason, it is critical to examine all possible data for every candidate, including the spectrum, before deciding on cluster membership.

 \subsection{Alessi 21}
 
 \begin{figure}[tb]
    \centering
    \includegraphics[width=0.99\textwidth]{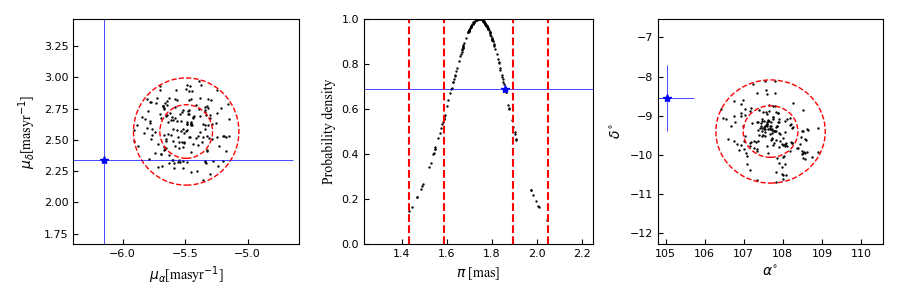}
    \caption{Astrometric data for Alessi 21. Proper motion (left), KDE of parallax values (centre) and equatorial position (right). Red dashed lines indicate 1, 2 sigma. Blue star with error bars indicates WD candidate.}
    \label{fig:alessi21}
\end{figure}\textbf{}

In Fig.~\ref{fig:alessi21} we present the astrometric data relating to the membership of the WD in the direction of the cluster Alessi~21. The WD's parallax is within 1$\sigma$ of the cluster mean value, while its proper motion and position with respect to the bulk of the cluster members puts it as an outlier in both these latter quantities. The WD is about the faintest object in Alessi~21 with a measurable proper motion, so while it does not make our cut for membership based on proper motion, its error is large (Fig.~\ref{fig:alessi21}, left panel). In addition, it is fairly common that cluster WDs do not generally lie near their respective clusters' centre. This may be the result of a small natal kick given to the newly formed WDs \citep{2003ApJ...595L..53F}. 

The cluster CMD, Fig.~\ref{fig:3panel1} (second panel from the left), is that of a young cluster with an age of $\sim$60$\pm20$ Myrs. This age is twice as old as the WEBDA age of 30 Myrs.  
The photometric mass estimate for the WD from the CMD is $\sim$1.0~M$_\odot$ if it is a cluster member. 
The WD cooling time to reach the observed magnitude (M$_G$ = 11.33) and colour ((Bp $-$ Rp)$_0$ = -0.40) is 121 Myrs. This cooling age is considerably longer than the age of the cluster suggesting that the WD is not a member. 
More information on the WD's properties are contained in Figure 13  where the Gemini spectrum of the WD is displayed. Fitting spectral WD models to this object yields log g = 8.36$\pm0.03$ and an effective temperature of 23,500$\pm200$~K. This is well-modeled by a WD of mass 0.85~M$_\odot$ with a cooling age of 70 Myrs. A WD of this mass is not entirely consistent with the photometric mass as can be seen in Fig.~\ref{fig:3panel1}, but it still lies well within its 1$\sigma$ error bars.

Even a cooling age for the WD of 70 Myrs is longer than the cluster age, leaving no time for main sequence evolution. We conclude from this discussion that the WD in Alessi~21 is unlikely a member of the cluster.

\subsection{ASCC 47}
 
The ASCC 47 WD can be considered a cluster member with a high degree of confidence. The astrometric data exhibited here together with our earlier publication on this WD and its host cluster \citep[][Fig.~\ref{fig:3panel1}]{2020ApJ...901L..14C}, make a very strong case for membership. The astrometric data provide the strongest case for membership of any WD found in our survey with all properties lying within about 1$\sigma$ of the mean cluster values. The cluster age from the CMD is 90$\pm20$ Myrs \citep{2020ApJ...901L..14C} and the photometry suggests a WD that is extremely hot and luminous. As we showed in \citeauthor{2020ApJ...901L..14C}, the WD is a hot magnetic DA, in fact the hottest and youngest WD known in any open star cluster. Its temperature is estimated to be 110,000~K and it has been cooling for only 250~kyrs. Analysis of the Zeeman splitting in the H-$\alpha$ line suggests that the WD is threaded by a magnetic field of about 1~MG on its surface.

\begin{figure}[tb]
    \centering
    \includegraphics[width=0.99\textwidth]{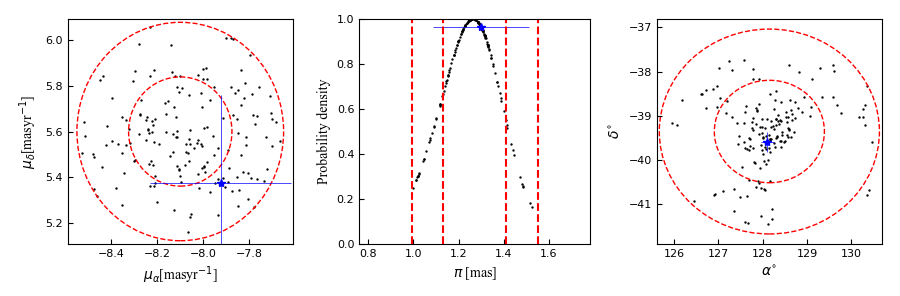}
    \caption{Astrometric data for ASCC 47 and its WD. Proper motion (left), KDE of parallax values (centre) and equatorial position (right). Red dashed lines indicate 1, 2 sigma. Blue star with error bars indicates WD candidate.}
    \label{fig:ascc47}
\end{figure}\textbf{}

The spectroscopic fit that we show in Fig.~\ref{fig:lines} employs non-magnetic models for a polluted hydrogen atmosphere developed by \citet{2010ApJ...720..581G}. The use of polluted models is crucial for such a hot white dwarf because the levitation of metals in the atmosphere due to high radiation pressure modifies the shape of the Balmer lines. However, we showed in \citeauthor{2020ApJ...901L..14C} that the values obtained from the fit of non-magnetic models is not reliable, especially for the value of the surface gravity, because when the magnetic field strength is close to 1~MG, gravity and magnetic field are degenerate in broadening the lines. In Table~\ref{tab:spec} we therefore report also the photometric value of $\log g$, which is the one we use in the determination of the mass of this WD.

\subsection{Alessi 8 and NGC 3114}

\begin{figure}[tb]
    \centering
    \includegraphics[width=0.99\textwidth]{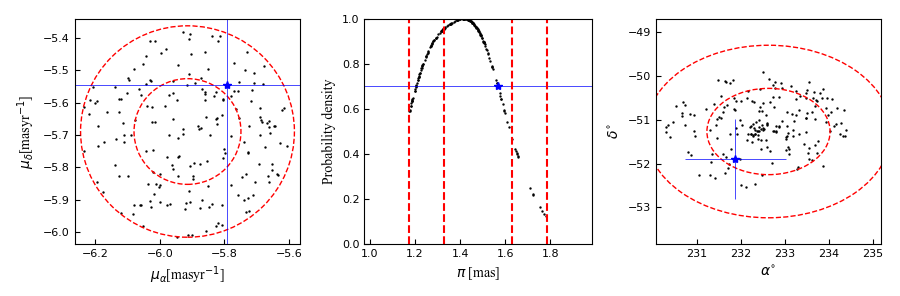}
    \caption{Astrometric data for Alessi 8. Proper motion (left), KDE of parallax values (centre) and equatorial position (right). Red dashed lines indicate 1 and 2~$\sigma$ limits. Blue star indicates WD candidate with its associated error bars.}
    \label{fig:alessi8}
\end{figure}

Alessi 8 appeared to be a promising WD cluster candidate. All its astrometric parameters lay within about 1$\sigma$ of the cluster mean properties (Fig.~\ref{fig:alessi8}) while its CMD suggested a cluster age near 100 Myrs and a massive WD $\sim$1.0~M$_\odot$.

However, a Gemini GMOS spectrum indicated that the object was not a WD. Similar comments apply to NGC 3114. In this case, the WD parallax agreed well with that of the cluster but it was somewhat of an outlier in proper motion and position. Here, the cluster age was estimated at $\sim$130 Myrs with the WD photometry implying a WD mass just below 1.0~M$_\odot$, but in this case the errors in the photometry were very large (see Table~\ref{tableclusters-all}). Again, a Gemini spectrum indicated the object was not a WD. We do not discuss these two stars any further here.

\textbf{}

 \subsection{Messier 47 (NGC 2422)}
 
 \begin{figure}[tb]
    \centering
    \includegraphics[width=0.99\textwidth]{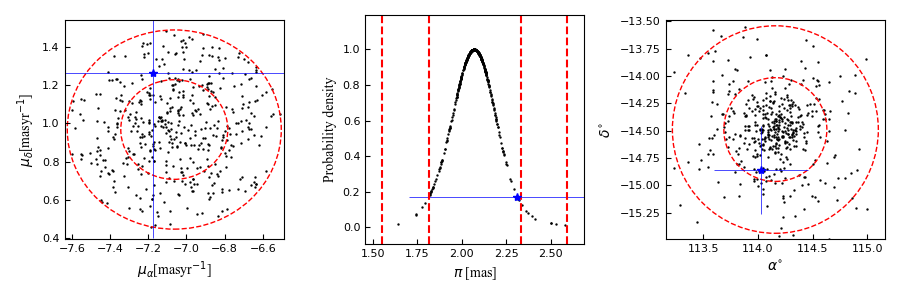}
    \caption{Astrometric data for M47 (NGC 2422). Proper motion (left), KDE of parallax values (centre) and equatorial position (right). Red dashed lines indicate 1, 2 sigma. Blue star indicates WD candidate.}
    \label{fig:m47}
\end{figure}\textbf{}

The WD in M 47 was discussed by our group in an earlier paper \citep{2019ApJ...880...75R}. The new analysis of the astrometric data presented here confirm the strong case for membership of the WD in the cluster. All the key astrometric parameters of the WD are within about 1$\sigma$ of the mean cluster values. The cluster itself is moderately rich and has an age, based on the isochrone fit in Fig.~\ref{fig:3panel1}, of about 150 Myrs.

The spectrum of this WD was discussed in detail in \cite{2019ApJ...880...75R}. It shows that the WD is a magnetic DB (helium atmosphere) with a mass of 1.06~M$_\odot$. Analysis of the Zeeman splitting in the helium lines indicates a magnetic field of about 2.5~MG. The WD cooling models displayed in Fig.~\ref{fig:3panel1} for this cluster are for DB stars and they seem to suggest that the WD has only a moderate mass of slightly under 0.8~M$_\odot$. The reason for this discrepancy is that, as we showed in our earlier paper using more extensive photometry than just that of Gaia (VPHAS+ Pan-STARRS), the WD has an infrared excess. The rather red Gaia colours are sensitive to this excess whereas in our original paper we established the photometric mass of the WD using VPHAS+ $u-$ and $g-$band data which are much less sensitive to the red excess.

\subsection{Stock 2}

The WD seen in the direction of Stock 2 has a high probability of being a cluster member with all its astrometric parameters lying within about 1$\sigma$ of that of the mean cluster values. This is quite a rich cluster, one of the two richest (together with M~47) of all those in which we identified a new massive WD member in this survey. 

\begin{figure}[tb]
    \centering
    \includegraphics[width=0.99\textwidth]{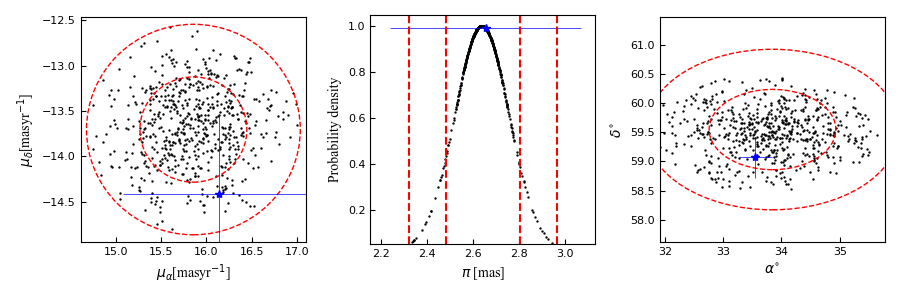}
    \caption{Astrometric data for Stock 2. Proper motion (left), KDE of parallax values (centre) and equatorial position (right). Red dashed lines indicate 1, 2 sigma. Blue star indicates WD candidate.}
    \label{fig:stock2}
\end{figure}\textbf{}

The age of the cluster from its CMD is quite uncertain. As can be seen in Fig.~\ref{fig:3panel2} (left panel), there is an enormous amount of scatter just below the turnoff region. It is not obvious what the source of this scatter is; it may be differential reddening, rotation of upper main sequence stars or binarity. We have established the cluster age from the three giant stars which together suggest a cluster age of 225$\pm50$ Myrs. This can be compared with the WEBDA value of 170 Myrs.

Our survey actually located 2 WDs in the direction of Stock 2, as can be seen in Fig.~\ref{fig:3panel2}). The cooler and fainter one of these seemed to be too low in mass to be a cluster member, so we did not pursue it further. It is a likely foreground object. In fact, the parallax for this star seems to confirm somewhat this suggestion, implying that it is about 10\% closer than the cluster.

The cluster CMD suggests a photometric mass for the WD of just under 1~M$_\odot$ if it is a cluster member. The color, (Bp $-$ Rp)$_0= -0.46$ , suggests a temperature of 29,000~K and together with the absolute magnitude (M$_G= 10.98$ ) implies that the time it took the WD to cool to these values is 74 Myrs (\citet{2010ApJ...720..581G}). Subtracting this from the cluster age of 225 Myrs leaves 151 Myrs for the main sequence lifetime of the WD precursor.  A star with this hydrogen burning lifetime has a mass of 4.5$\pm0.2$~M$_\odot$, which would be the mass of the precursor of this WD. 
 
Analysis of the spectrum of the WD yields $\log g = 8.58\pm0.05$ and an effective temperature of $25,400\pm300$~K. This is well-modeled by a WD of mass 0.99~M$_\odot$ with a cooling age of 110 Myrs. A WD of this mass is entirely consistent with the photometric mass as can be seen in Fig.~\ref{fig:3panel2}. A cooling age for the WD of 110 Myrs implies that the progenitor spent 115 million years on the main sequence in this 225 Myr cluster. A star of mass 5.0~M$_\odot$ spends this amount of time in the main sequence phase.  
Since the spectroscopic results are expected to be more precise, we adopt these for the properties of the WD in Stock 2.  

\subsection{ASCC 113} 

The Gaia parallax of the WD in the direction of ASCC 113 is within 1$\sigma$ of the mean cluster value while the proper motion and location on the sky put it as a slight outlier. The cluster CMD seen in Fig.~\ref{fig:3panel2}, second panel from the left, implies a cluster age of about 240$\pm40$ Myrs and suggests a photometric mass for the WD near 1.0~M$_\odot$.
 
\begin{figure}[tb]
    \centering
    \includegraphics[width=0.99\textwidth]{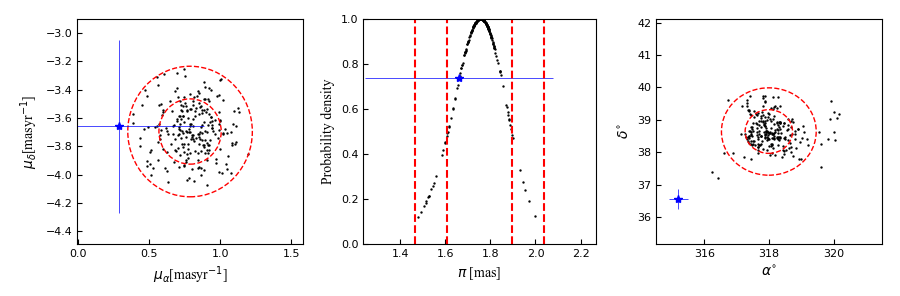}
    \caption{Astrometric data for ASCC 113. Proper motion (left), KDE of parallax values (centre) and equatorial position (right). Red dashed lines indicate 1, 2 sigma. Blue star indicates WD candidate.}
    \label{fig:ascc113}
\end{figure}\textbf{}

Fig. 12 displays our full Gemini-North spectrum of the ASCC 113 WD while Fig. 13 shows the best spectral fits to the four Balmer lines H$\beta$ through to H$\epsilon$. The stellar parameters derived from these fits are $\log g = 8.71\pm0.07$ and $T = 25,400\pm300$~K.
These WD properties are well-modeled by a WD of mass 1.06~M$_\odot$ with a cooling age of 160 Myrs. This WD mass, as can be seen in Fig.~\ref{fig:3panel2} is also consistent with the photometry of the WD given the photometric errors. This cooling age, coupled to the age of the cluster yields the WD precursor's main sequence lifetime: about 80 Myrs. From the Padova stellar evolution models \citep{2012MNRAS.427..127B}, we find that a star of 5.8~M$_\odot$ spends 80 Myrs on the main sequence. 

The ASCC~113 WD is thus tied with the WD in M~47 as the most massive WD we found in our survey for massive cluster WDs. The M47 WD is, however, quite a different WD, being a magnetic DB star.  

\subsection{Messier 39 (NGC 7092)}

The astrometric data for the WD in Messier 39 present a reasonably strong case for its membership in the cluster. The parallax is well within 1$\sigma$ of the cluster mean, the position on the sky is within 2$\sigma$ while the proper motion is just outside 2$\sigma$.

The CMD of this cluster, Fig.~\ref{fig:3panel2}, suggests an age near 280$\pm20$ Myrs and yields a photometric mass for its WD near 0.9~M$_\odot$.  The spectrum of the WD was discussed in detail in \citet{2020ApJ...901L..14C} and yielded $\log g = 8.87\pm0.07$ and a temperature of $18,400\pm300$~K. This apparently high gravity is partially due to the presence of a magnetic field which has the effect of broadening the spectral lines, hence mimicking a higher mass WD. The photometric $\log g$, which we use in this case as it is largely unaffected by the presence of the magnetic field, suggested a value of $8.54\pm0.04$ for $\log g$ and together with the temperature yielded a mass of 0.95$\pm0.02$~M$_\odot$.
\begin{figure}[tb]
    \centering
    \includegraphics[width=0.99\textwidth]{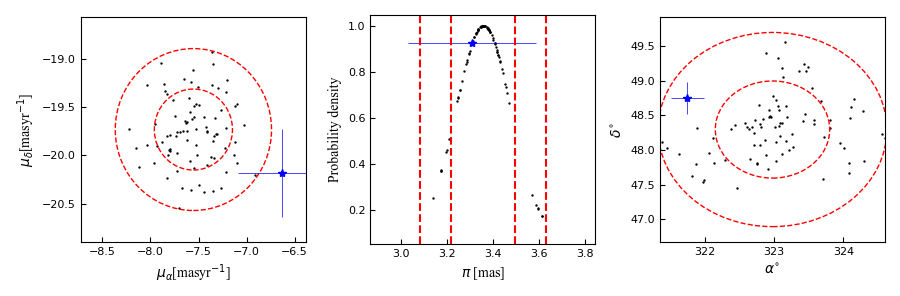}
    \caption{Astrometric data for M 39 (NGC 7092). Proper motion (left), KDE of parallax values (centre) and equatorial position (right). Red dashed lines indicate 1, 2 sigma. Blue star with error bars indicates WD candidate.}
    \label{fig:m39}
\end{figure}\textbf{}

\subsection {Stock 12} 

The astrometric data for the WD in Stock 12 places all the parameters at just about the 2$\sigma$-level with respect to the mean cluster values. The Gaia CMD Gaia (Fig.~\ref{fig:3panel2}) suggests a rather low mass for the WD with our best estimate being 0.47~M$_\odot$ while the temperature from this photometry is about 18,000~K.

\begin{figure}[tb]
    \centering
    \includegraphics[width=0.99\textwidth]{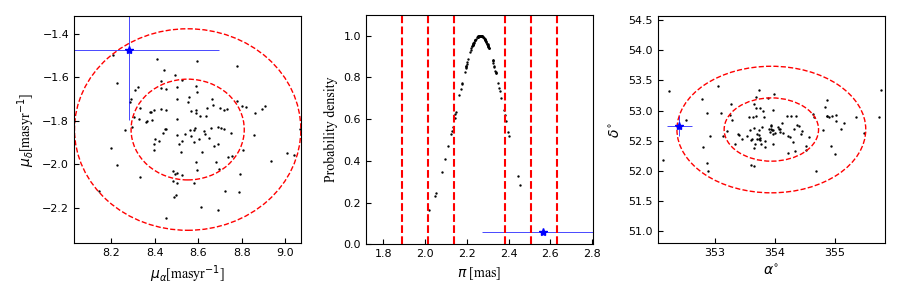}
    \caption{Astrometric data for Stock 12. Proper motion (left), KDE of parallax values (centre) and equatorial position (right). Red dashed lines indicate 1, 2 sigma. Blue star indicates WD candidate.}
    \label{fig:stock12}
\end{figure}\textbf{}

The spectroscopic values determined from the Gemini North spectrum are in strong disagreement with the photometric values for the mass and temperature. The measured spectroscopic value of $\log g$ is 8.50 with a temperature of about 31,600~K suggesting a much higher mass (0.94~M$_\odot$) and hotter WD. Unlike the WDs in M~39 and ASCC~47, there is no obvious evidence of a magnetic field in the Stock~12 spectrum, so it does not appear that the lines have the excess broadening that could account for this higher log g value. This possibility cannot be entirely excluded however. In the Montreal WD database \citep{2017ASPC..509....3D} the star is listed as a DA WD (32,158~K, $\log g$ 8.30, 0.83~M$_\odot$). These numbers come from analysis of an SDSS spectrum and are in reasonable agreement with our spectroscopic results except that our log g value, which we adopt, is larger. The Sloan colours for this WD ($(u-g)_0 = -0.21$ and $(g - r)_0 = -0.38$) also suggest a temperature (25,000~K) significantly cooler than the spectroscopic values. Perhaps the reddening has been underestimated. We will adopt the Gemini spectroscopic parameters for this WD in order to estimate its cooling time and precursor mass. 

The time for a 0.94~M$_\odot$ WD to cool to its current observed temperature of 31,600~K is only  $\sim$37 Myrs. Since we estimate the cluster age from the CMD Fig.~\ref{fig:3panel2} to be 300$\pm50$ Myrs, the main sequence lifetime of the progenitor was $\sim$263 Myrs implying that the WD precursor was a star of mass 3.7$\pm0.4$~M$_\odot$. While not a particularly massive WD, this object is consistent with other objects in its location in the IFMR (Fig. 16).

\begin{figure}
    \centering
    \includegraphics[width=\textwidth]{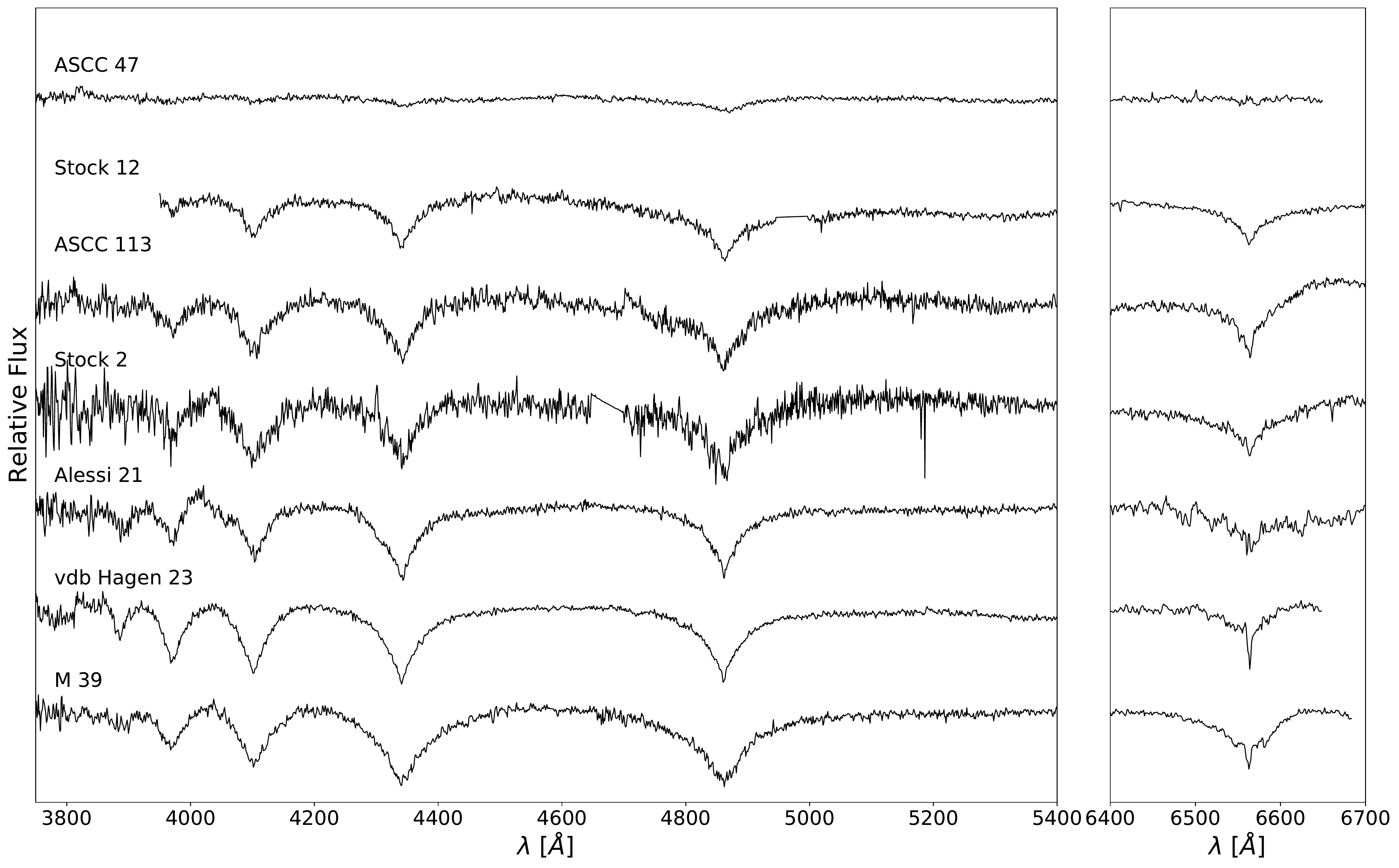}
    \includegraphics[width=0.97\textwidth]{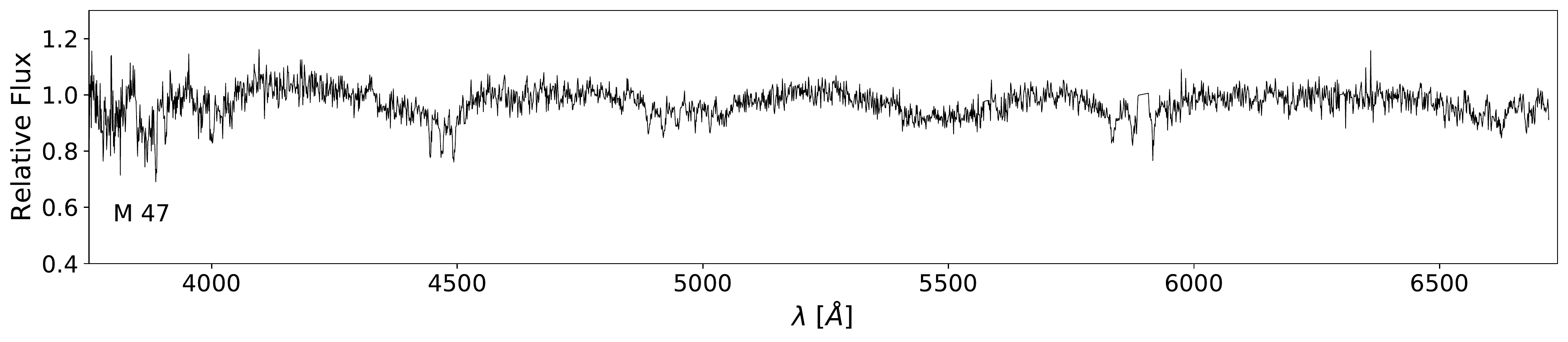}
     
    \caption{Spectra of the cluster WDs obtained with Gemini. The upper panels show the DA WDs ordered from the hottest (ASCC 47) to the coldest (M 39). The upper-right panel shows H$\alpha$ for each of the spectra. The lower panel shows the only DB white dwarf found in this survey (M 47). } 
    \label{fig:totspec}
\end{figure}
  
\begin{figure}
    \centering
    \includegraphics[width=\textwidth]{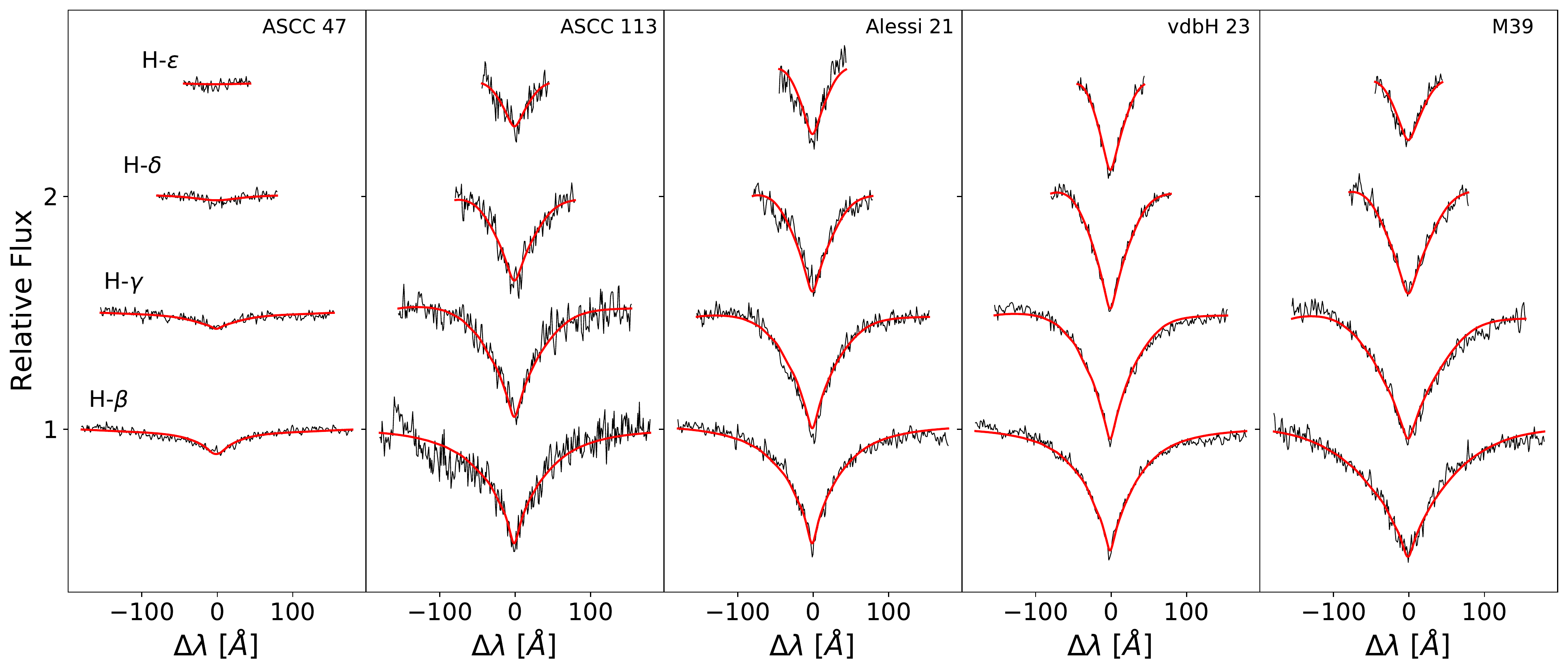}
    \includegraphics[width=0.44\textwidth]{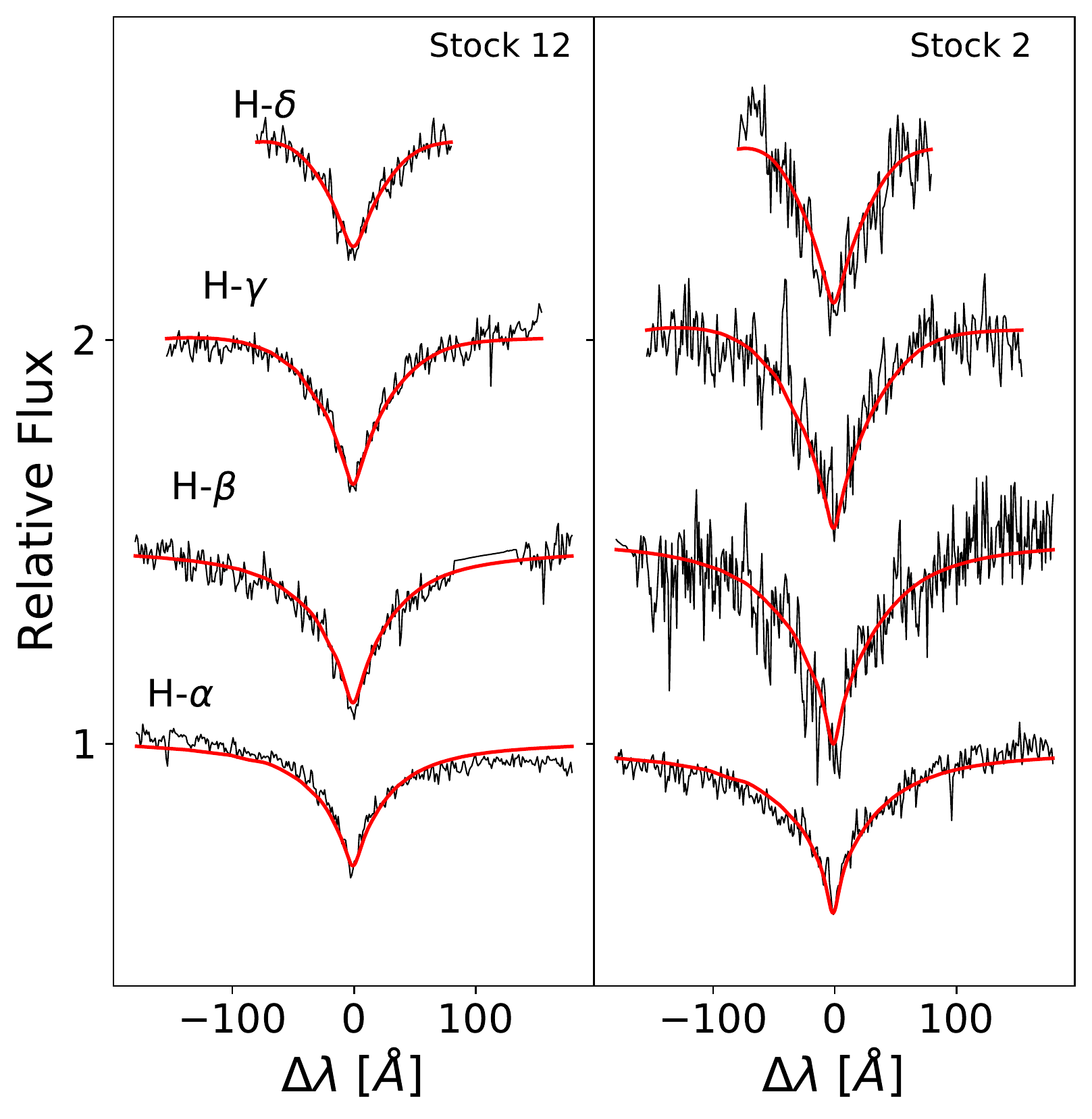}
    \caption{Fit to the Balmer lines for the cluster WDs. This was used to obtain an estimate of the surface gravity and effective temperature for each WD. For the WDs in the upper panel, the Balmer lines H$\beta$ to H$\epsilon$ were used in the fit, while for the WDs in the lower panel, the Balmer lines H$\alpha$ to H$\delta$ were used to determine the stellar parameters. In each panel the best fit is shown in red, while the values for the best fits are given in Table~\ref{tab:spec}. } 
    \label{fig:lines}
\end{figure}

\begin{figure}[h]
    \centering
    \includegraphics[width=0.9\textwidth,clip,trim=0 2in 0 0 ]{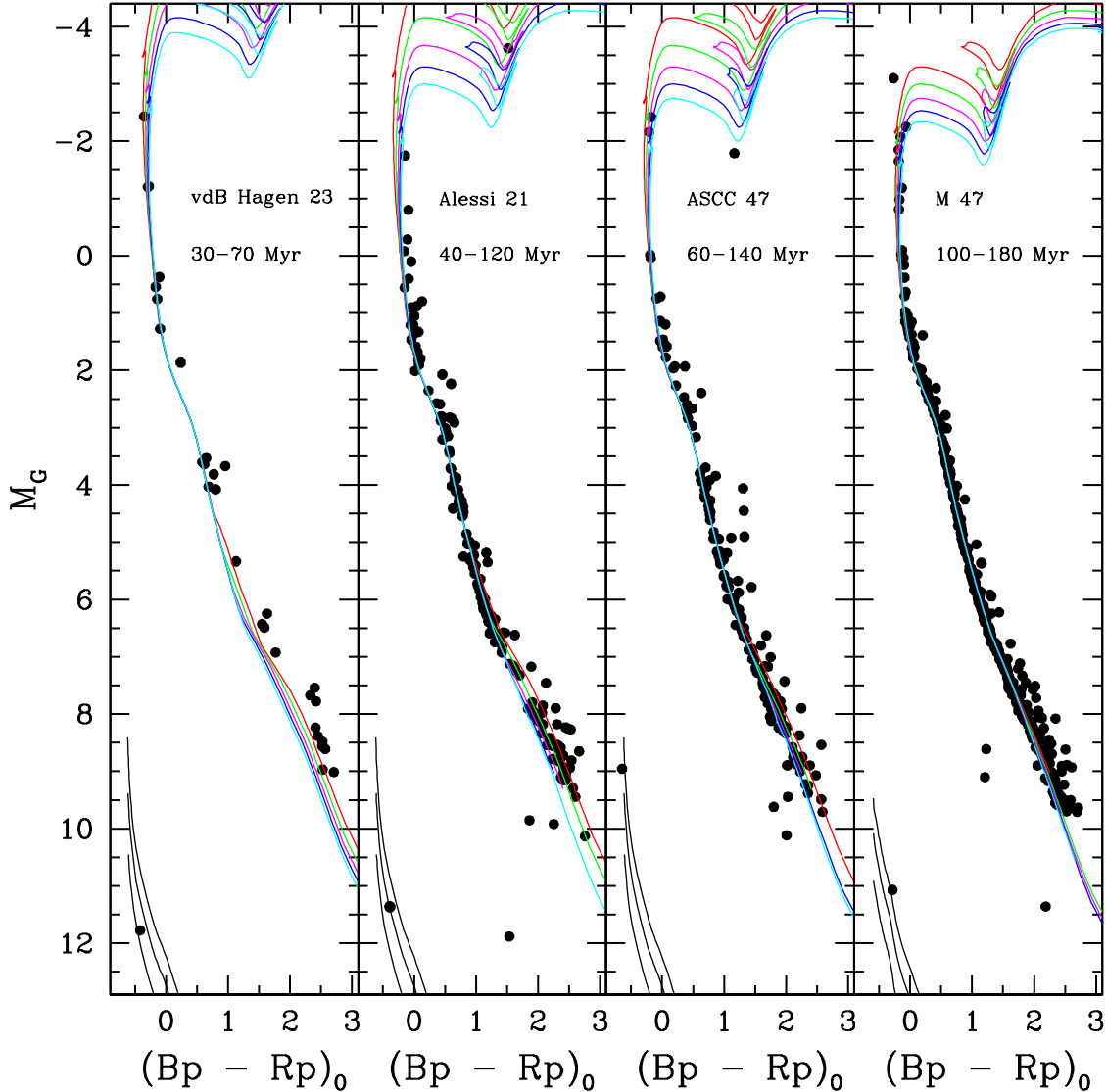}
    \caption{A four panel diagram showing the CMDs (increasing in age) for the clusters vdB Hagen 23, Alessi 21 ASCC 47 and  Messier 47 (NGC 2422). DA WD cooling sequences \citep{2020ApJ...901...93B}  for 0.8, 1.0 and 1.2 M$_\odot$ are shown for all the clusters except M 47 which are DB sequences of the same mass as this WD is a known DB star (Richer et al. 2019). The WD does not appear as massive here as we claimed in (Richer 2019) as its has an IR-excess which is important in the Gaia filters and less so in the u, g VPHAS+ filters which we used in the previous paper. Isochrone ages are indicated in the diagram.}
    \label{fig:3panel1}
\end{figure}
 
\begin{figure}[htb]
    \centering
    \includegraphics[width=0.9\textwidth,clip,trim=0 2in 0 0]{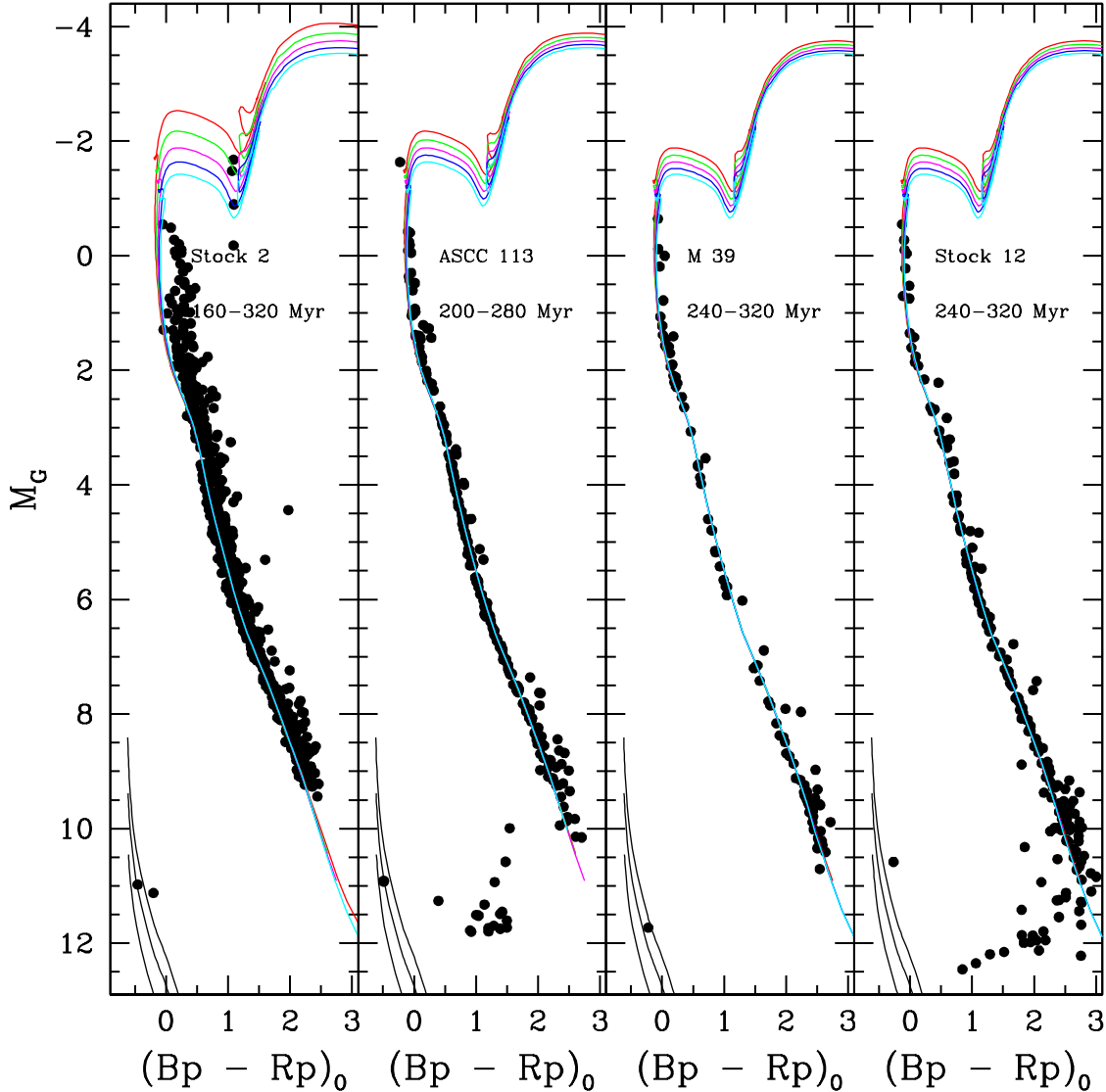}
    \caption{A four panel diagram showing the CMDs (increasing in age) for the clusters Stock 2, ASCC 113, Messier 39 (NGC 7092) and Stock 12. DA cooling sequences  \citep{2020ApJ...901...93B}
        for 0.8, 1.0 and 1.2 M$_\odot$ are included. Only the bluer and brighter WD candidate in Stock 2 is discussed here as the other star does not appear to be a likely cluster member given its potential mass. Isochrone ages are indicated in the diagram.}
    \label{fig:3panel2}
\end{figure}

\begin{deluxetable*}{lcccc}[tb]
\tablecaption{Properties of New Massive Cluster White Dwarf Candidates\label{tableclusters2}}
\tablewidth{700pt}
\tabletypesize{\scriptsize}
\tablehead{
\colhead{Cluster Name} &  
\colhead{Cluster Age (Myrs)} &  \colhead {WD Mass ($M{_\odot}$)} & \colhead {Precursor Mass ($M{_\odot}$)} &  \colhead {Comments}
} 
\startdata
vdB-Hagen~23 &  55$\pm20$ & 0.63$\pm0.03$ & .......  & unlikely cluster member \\
Alessi 21 &  60$\pm20$ & 0.85$\pm0.02$& ....... & unlikely cluster member \\
ASCC 47  & 90$\pm20$  &1.01$\pm0.02$ & 5.6$\pm0.8$& magnetic DA, hottest cluster WD known\\
Alessi 8 &  100$\pm30$ & ....... & ....... & not a WD \\
NGC 3114 & 130$\pm25$ &....... & .......& not a WD\\
M 47 &  150$\pm20$ &1.06$\pm0.05$ & 6.1$\pm0.5$& magnetic DB, IR excess\\
Stock 2 & 225$\pm50$ & 0.99$\pm0.03$ & 5.5$\pm1.5$& cluster age uncertain \\
ASCC 113 & 240$\pm40$ & 1.06$\pm0.10$ & 5.8$\pm0.8$ & tied (M47) as most massive WD found in this survey \\
M 39 & 280$\pm20$ &0.95$\pm0.02$ &5.4$\pm0.6$ & magnetic DA \\
Stock 12 & 300$\pm50$ & 0.94$\pm0.02$ & 3.7$\pm0.4$&  magnetic DA? \\
\enddata
\end{deluxetable*}

\begin{deluxetable*}{lcccc}[tb]
\tablecaption{WD Parameters from the Spectroscopic Fits\label{tab:spec}}
\tablewidth{700pt}
\tabletypesize{\scriptsize}
\tablehead{
\colhead{Cluster Name} &  
\colhead{$\log~g~[\log(\rm{cm}~\rm{s}^{-2})]$} &  \colhead {Effective Temperature [K]} &  \colhead {Comments}
} 
\startdata
vdB-Hagen~23 &  8.02$\pm$0.03 & 19,600$\pm$200 & ....... \\
Alessi 21 &  8.36$\pm$0.03 & 23,500$\pm$200 & .......\\
ASCC 47  & 8.93$\pm$0.13  &114,000$\pm$3,000 & \makecell{$\log g$ affected by magnetic field, \\ photometric $\log g =8.47\pm0.05$ (see Caiazzo et al. 2020)} \\
Stock 2 & 8.58$\pm$0.05 & 24,900$\pm$400 & photometric temperature 29,000~K  \\
ASCC 113 & 8.71$\pm$0.07 & 25,400$\pm$300  & ....... \\
M39 & 8.93$\pm$0.06 & 18,100$\pm$300 & \makecell{$\log g$ affected by magnetic field, \\ photometric $\log g =8.54\pm0.04$ (see Caiazzo et al. 2020)} \\
Stock 12 & 8.50$\pm$0.04 & 31,600$\pm$200 &  estimate of $\log g$ could be affected by magnetic field \\
\enddata
\end{deluxetable*}


\section{Cluster White Dwarfs Located Far From Their Birthplace }

In section~\ref{sec:expectedNwd}, we calculated the expected number of white dwarfs in each cluster, $\langle{N_{WD}}\rangle$. Although the error in $\langle{N_{WD}}\rangle$ is large for some clusters, $\langle{N_{WD}}\rangle$ in many clusters exceeds the observed number of WDs, providing evidence for a deficiency of observed WDs. Out of the 163 clusters for which we were able to fit isochrones, $\langle{N_{WD}}\rangle > 1$ for $\sim 65\%$  of the clusters analyzed. This suggests that there may exist a widespread phenomena of missing WDs in many young, open clusters. 

This discrepancy is somewhat mitigated by the fact that, for the most distant clusters, we did not expect to detect the full population $\langle{N_{WD}}\rangle$ of white dwarfs because the more massive are now cold and dim, below the detectability threshold for Gaia. Another potential factor is that WDs may be given a modest kick (a few km/sec) when born \citep{2003ApJ...595L..53F}. As a result, as the WDs age and move away from the center of the cluster, they may then often be found near or beyond the periphery of the cluster in positional space and hence be excluded when considering data within a chosen radial distance. With even a kick of 1 km/sec, one of these WDs could travel around 100 pc away from the cluster in 100 Myr, making it imperative to consider potentially kicked WDs that we may have missed in our initial search.

In order to explore this possibility further, we selected 40 of our surveyed clusters and searched within a wide radius for potential stripped WDs. This subset of clusters has WEBDA distances below 800 pc and ages between 10 Myr and 150 Myr (corresponding to turnoff masses of 14.5 M$_\odot$ to 4.3 M$_\odot$). For each cluster, a query was made to the \cite{Gentile_Fusillo_2018} catalog for high probability WD candidates (their Pwd $\ge0.75$) within $20 \times$ the literature cluster radius and $2\sigma$ in proper motion of the cluster centre, accounting for the WD proper motion error provided by Gaia. Using such a large search radius for these young clusters should allow us to capture most of the expelled WDs. After removing the objects that already appeared in our original narrow search, a total of 992 potential cluster member candidate WDs were identified in the 40 selected clusters. 

\begin{figure}
    \centering
    \includegraphics[width=0.8\textwidth]{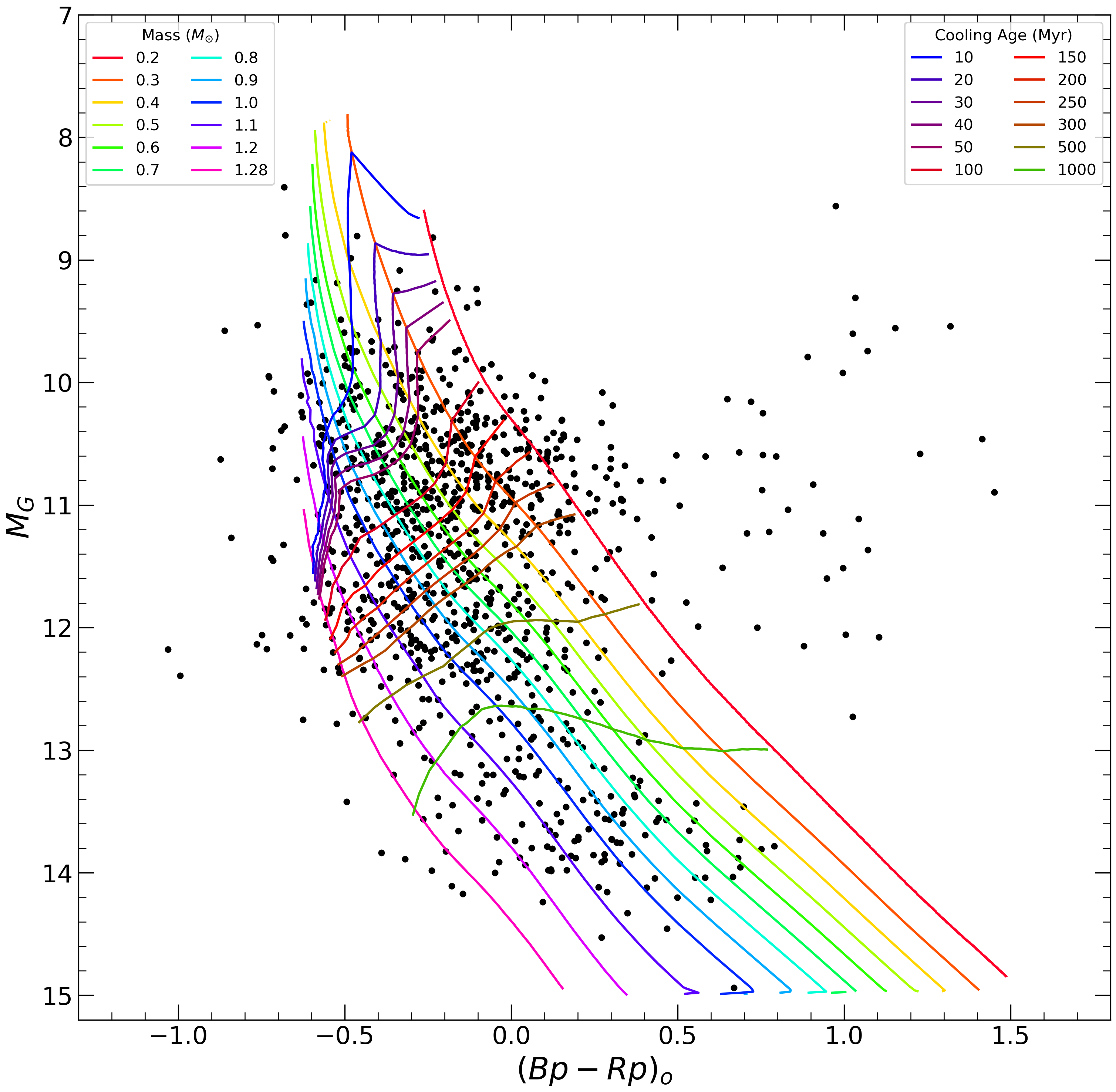}
        \caption{Candidate cluster member WDs identified in the wide search. WDs in the CMD have been dereddened using their associated cluster values. Primarily vertical tracks show mass models from $0.20$ M$_\odot$ (far right) to $1.28$ M$_\odot$ (far left), while primarily horizontal ones show cooling ages from $10$ Myr (top) to $1$ Gyr (bottom).} 
    \label{fig:widesearch}
\end{figure}

Candidate WDs were dereddened using associated cluster parameters, allowing their mass and cooling ages to be estimated using the python package \texttt{WD\_models}\footnote{https://github.com/SihaoCheng/WD$\_$models}. WD masses $\leq1.0$ M$_\odot$ are modeled using the Montreal group CO cooling models \citep{B_dard_2020}\footnote{http://www.astro.umontreal.ca/$\sim$bergeron/CoolingModels/}, while those $>1.0$ M$_\odot$ use O/Ne core models \citep{2019A&A...625A..87C}\footnote{http://evolgroup.fcaglp.unlp.edu.ar/TRACKS/ultramassive.html}, up to a maximum of $1.28$ M$_\odot$. Progenitor masses were estimated for each WD using the IFMR from \citet{2018ApJ...866...21C}.
Fig.~\ref{fig:widesearch} shows the full set of candidate WDs, along with mass models for WD masses between $0.20$ M$_\odot$ and $1.28$ M$_\odot$, for cooling ages between $10$ Myr and $1$ Gyr.  

To focus on the most likely cluster members, only WDs whose upper bound cooling age was younger than the cluster age plus the $1~\sigma$ error were retained (see Table~\ref{tableclusters}). Additionally, WDs whose progenitor mass $+3\sigma$ was less than the cluster main sequence turnoff were removed. This loose cutoff allowed us to remove some clear outliers without excluding candidates that may be reasonable given their location in the cluster CMD. There were a handful of objects with masses outside the model grid range ($>1.28$ M$_\odot$) for which we did not have cooling ages. For these candidates, their positions in the cluster CMDs were examined individually for candidacy. In the end, of the 992 WDs in the wide search, 151 of them were retained as possible cluster members. The candidates' physical parameters are shown in Table~\ref{widesearchtable}, along with the associated cluster and its WEBDA diameter. For additional cluster parameters see Table~\ref{tableclusters}. 

We stress that these additional 151 candidate WDs require significant follow-up, including spectroscopy and a detailed analysis of their proper motions, to better assess their candidacy and that they do not represent bona-fide cluster members. For these same 40 clusters, all the white dwarfs should still be visible in Gaia and the number of expected white dwarfs is $\langle{N_{WD}}\rangle = 203 \pm 39$, which is within $2\sigma$ of the wide search survey results. Although this result is encouraging, we expect a large fraction of the candidates not to be cluster members. Further studies investigating the physical mechanism behind WD kicks or modelling cluster diffusion in detail could be key in determining where the majority of WDs in open clusters may be hiding. The results of the wide search will be examined in more detail in future work.


 \section{Implications of the Search for Massive Cluster WDs}
 
From the discussion in the previous sections, it is clear that we expected to find many more massive WDs than we ultimately located. As hinted at earlier, this may be a result of a kick imparted to the WD when born. Though the wide search potentially located some of these objects, further examination would be required to verify membership in their respective clusters. Even if all these WDs were bona-fide cluster members (which is surely not correct), many massive cluster WDs remain unaccounted for.
 
In Fig. 17 we present a new version of the IFMR concentrating on the high mass end of the relationship. In this plot we only include those cluster WDs that appear in Gaia, pass all our astrometric tests for cluster membership and have an initial mass in excess of 2.5~M${_\odot}$. A number of WDs that appeared in previous instances of this plot \cite[e.g.][]{2018ApJ...866...21C} were eliminated from this diagram simply because they are too faint to be in Gaia (e.g. several WDs in NGC 2168, NGC 2323-WD10). Others did not pass all our astrometric tests (e.g. NGC 3532-J1107-584). 
 
\begin{figure}[h]
    \centering
    \includegraphics[width=0.99\textwidth,clip,trim=0 2in 0 0]{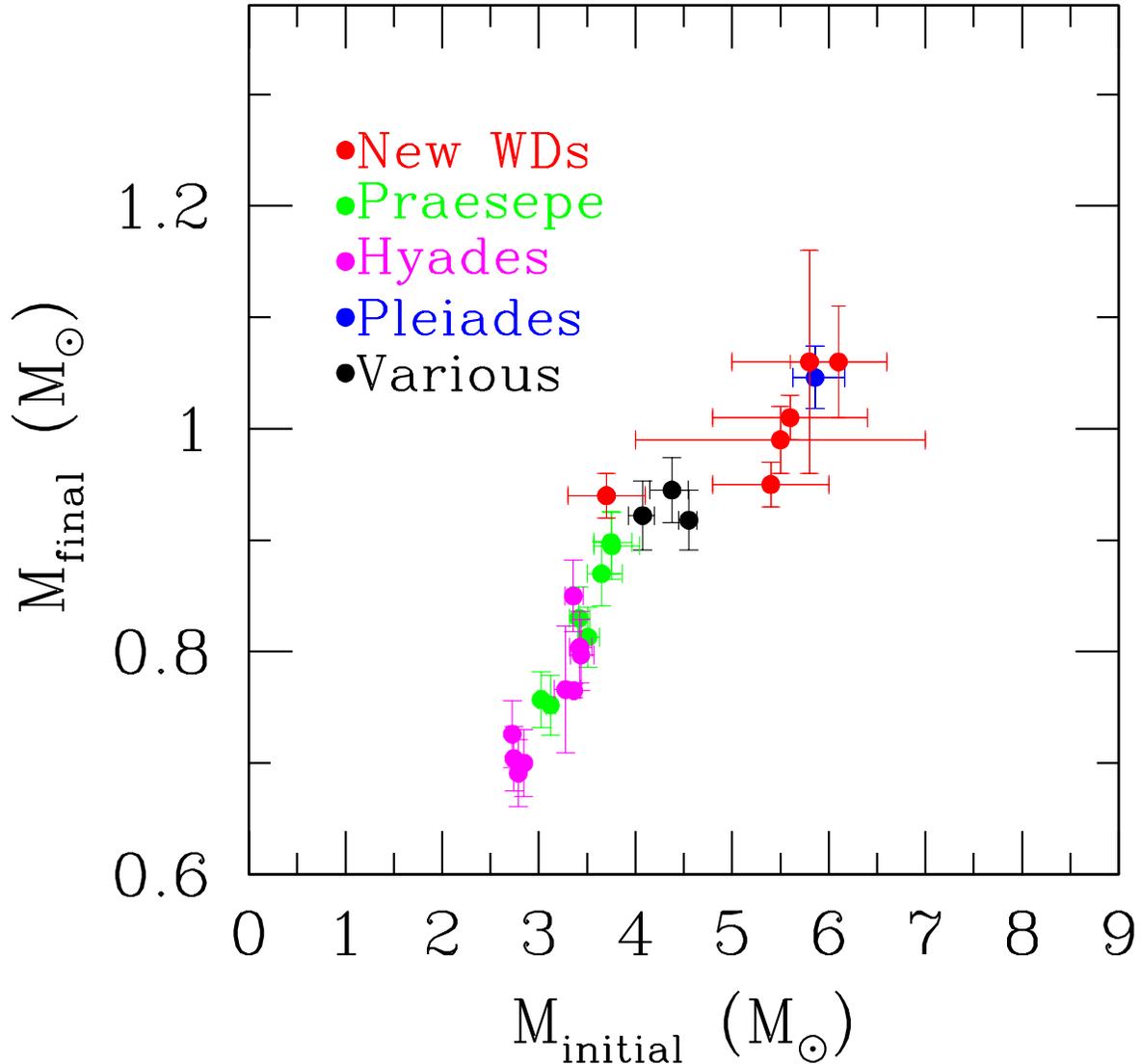}
    \caption{IFMR for stars with initial masses in excess of $\sim2.5~M{_\odot}$ including the new cluster WDs discovered in the current survey plus those from the literature which pass all our astrometric tests. The stars plotted as ``Various" with black dots are NGC~3532-J1106-584, NGC~3532-J1106-590 and NGC~2516-5. Many of the stars appearing in earlier versions of this diagram are not present here as they did not pass our  strict astrometric requirements or  simply because they were too faint for Gaia. The Hyades WDs were not actually found in our survey because of the large angular extent of the cluster, but are included here nonetheless.  }
    \label{fig:ifmr}
\end{figure}

The current diagram has much less scatter than earlier versions. Of interest is that we were not able to locate WD precursors in these clusters that were more massive than about 6~M${_\odot}$. Except possibly for GD~50, there is no cluster WD known that passes strict astrometric tests and that evolved from a single main sequence star in excess of this mass. In fact, since GD~50 is not really a bona-fide member of any well-defined stellar cluster (it may be part of the AB Doradus moving group) and considering that it may even be a merger remnant \citep{1996ApJ...461L.103V}, it was excluded from the current version of the IFMR. The most massive WD precursor remaining in Fig. \ref{fig:ifmr} is the 6.1~M${_\odot}$ precursor to the magnetic DB in Messier 47. 
A mass of 6~M${_\odot}$ seems to be the upper limit to WD precursors that we found {\it in the clusters that we surveyed here}. This may be a fundamentally important result or it may be a consequence of massive WDs either quickly leaving their clusters after birth through some dynamical event or being undetectable due to being located in binary systems.  We found that the upper limit of the initial stellar mass for white dwarf formation only has a modest effect of about 30\% on the expected number of white dwarfs. On the other hand, the multiplicity fraction amongst high mass stars is thought to be substantial, even greater than 75\% \citep[for  recent discussions see][]{2007ApJ...670..747K,2017ApJS..230...15M}; making this is an important consideration.
 
If this low upper limit to WD formation is a result of stellar evolution and not a consequence of dynamics or binarity, it exacerbates the discrepancy between the observed SN II rate and the number of stars with masses in excess of that capable of producing such a SN. \cite{2011ApJ...738..154H} suggested that a rate that was twice as high as observed would occur if all stars more massive than 8~$M{_\odot}$ produce SN II. If this limit is reduced to 6~$M{_\odot}$ this discrepancy becomes at least a factor of three. Additionally, it appears that single stars do not leave behind electron-degenerate remnants that even remotely approach the Chandrasekhar Limit --- from our survey the highest mass WDs produced appear to be below 1.1~$M{_\odot}$.

%



\appendix
Table~\ref{widesearchtable} contains select parameters for clusters associated with wide search WDs, as well as a number of WD parameter estimates if they are indeed cluster members. We emphasize that this is not a list of WDs that are necessarily cluster members, instead is a list of WDs that could be part of these clusters and are worth considering in future work.

Table~\ref{tableclusters} contains information on the young open clusters that were searched for potential massive WDs in the narrow search. While WEBDA provided the source list for the majority of these clusters (with the exception of 6, as discussed in section 2), all the astrometric information contained in the table was derived by the authors using the Gaia DR2 database.
 
Table~\ref{tableWDs-excluded} contains a list of objects which appear in \cite{2018ApJ...866...21C} as potential cluster member WDs but do not meet our search criteria, along with corresponding Gaia data, if available. Any objects that appear in this reference and make our cuts have already been listed in Table~\ref{tableclusters-all}.
 

\bibliography{sample63.bib}
\bibliographystyle{aasjournal}


\acknowledgments
This work has made use of data from the European Space Agency (ESA) mission
{\it Gaia} (\url{https://www.cosmos.esa.int/gaia}), processed by the {\it Gaia}
Data Processing and Analysis Consortium (DPAC,
\url{https://www.cosmos.esa.int/web/gaia/dpac/consortium}). Funding for the DPAC
has been provided by national institutions, in particular the institutions
participating in the {\it Gaia} Multilateral Agreement.

\facilities{Gaia(DR2), Gemini North and South}


\software{Astropy \citep{2013A&A...558A..33A, 2018AJ....156..123A},  
          Cloudy \citep{2013RMxAA..49..137F}, 
          SExtractor \citep{1996A&AS..117..393B}
          } 


\startlongtable
\begin{deluxetable*}{ccccccccccccc}
\tablecaption{Wide Search WD Cluster Member Candidates\label{widesearchtable}}
\tablewidth{1100pt}
\tabletypesize{\scriptsize}
\allowdisplaybreaks
\setlength{\tabcolsep}{4pt}
\tablehead{
\\
\multicolumn{3}{c}{Cluster} 
& \multicolumn{10}{c}{WD} 
\\
\multicolumn{3}{c}{\rule{4.5cm}{.02cm}}
& \multicolumn{10}{c}{\rule{12.6cm}{.02cm}}
\\
\colhead{\hspace{.01cm}Name}\hspace{.01cm} 
& \colhead{\hspace{.01cm}Diam.}\hspace{.01cm} 
& \colhead{\hspace{.01cm}Turnoff Mass}\hspace{.01cm}
& \colhead{\hspace{.01cm}Gaia Source ID}\hspace{.1cm} 
& \colhead{\hspace{.01cm}G$_{0}$}\hspace{.01cm}
& \colhead{\hspace{.01cm}(Bp-Rp)$_{0}$}\hspace{.01cm}
& \colhead{\hspace{.01cm}Dist.}\hspace{.01cm} 
& \multicolumn{3}{c}{\hspace{.01cm}Mass}\hspace{.01cm}
& \multicolumn{3}{c}{\hspace{.01cm}Cooling Age}\hspace{.01cm}
\\
& [pc] 
& [$M_\odot$] 
& 
& [mag] 
& 
& [Cl$_{\rm rad}$] 
& \multicolumn{3}{c}{[$M_\odot$]} 
& \multicolumn{3}{c}{[Myr]}
\\
\multicolumn{7}{c}{}
& \multicolumn{3}{c}{\rule{2.7cm}{.02cm}}
& \multicolumn{3}{c}{\rule{3.1cm}{.02cm}}
\\
\multicolumn{7}{c}{}
& \colhead{Est.}
& \colhead{Upper}
& \colhead{Lower}
& \colhead{Est.}
& \colhead{Upper}
& \colhead{Lower}
\\
(1) & (2) & (3) & (4) & (5) & (6) & (7) & (8) & (9) & (10) & (11) & (12) & (13)
 } 
\startdata
Alessi 5      & 4.76  & 5.00 & 5352954771936845440 & 11.02 & -0.51 & 18.04 & 1.07               & $>$1.28 & 0.81               & 55.48                 & \ldots & 89.14                 \\
Alessi 8      & 7.32  & 4.64 & 5887666586717940224 & 10.07 & -0.48 & 16.18 & 0.69               & 0.93               & 0.53               & 10.55                 & 3.14                  & 18.92                 \\
              &       &      & 5982381370178315904 & 9.95  & -0.73 & 9.50  & $>$1.28 & $>$1.28 & 1.13               & \ldots & \ldots & 0.14                  \\
Alessi 12     & 9.45  & 4.64 & 1762939111272873216 & 11.08 & -0.58 & 16.10 & 1.20 & $>$1.28 & 1.06 & 6.55  & \ldots & 68.19  \\
              &       &      & 1805484267034684672 & 10.98 & -0.30 & 19.39 & 0.68 & 0.87    & 0.50 & 98.47 & 74.50  & 136.19 \\
              &       &      & 1817445136772265856 & 11.20 & -0.27 & 6.22  & 0.72 & 0.95               & 0.50               & 142.19                & 101.40                & 198.14                \\
              &       &      & 1832754079548398848 & 11.22 & -0.35 & 10.54 & 0.85               & 1.00               & 0.70               & 120.30                & 98.30                 & 150.07                \\
              &       &      & 1840424272662732544 & 11.36 & -0.20 & 16.57 & 0.67               & 0.94               & 0.46               & 203.37                & 136.24                & 290.33                \\
\enddata 
{\medskip {\bf Notes.} Selected parameters for those clusters associated with wide search WDs. Also included are WD parameter estimates assuming they are cluster members. Columns: (1) cluster name; (2) diameter (from WEBDA); (3) current cluster main sequence turnoff mass (using IFMR from \cite{2018ApJ...866...21C}); (4) Gaia DR2 WD source ID \citep{gaia_2018}; (5) WD absolute G magnitude (using associated cluster distance and reddening and $A_G = 2.059 E(Bp - Rp)$ , see Table.~\ref{tableclusters}); (6) intrinsic ($Bp-Rp$) color (similarly dereddened); (7) distance from cluster center in units of cluster radius; (8) mass estimate WD from cooling model fits; (9-10) $1\sigma$ upper and lower bound estimates on WD mass; (11) WD cooling age estimate from model fits; (12-13) $1\sigma$ upper and lower bound estimates on cooling age.}
\end{deluxetable*}

\begin{longrotatetable}
\begin{deluxetable*}{ccccccccccccccccc}
\tablecaption{Clusters studied\label{tableclusters}}
\tabletypesize{\scriptsize}
\allowdisplaybreaks

\tablehead{
& & & \multicolumn{4}{c}{Position}  & \multicolumn{4}{c}{Proper Motion} & \multicolumn{2}{c}{Parallax} &  \multicolumn{3}{c}{$E\rm{(Bp-Rp)}$} & Expected  \\
\colhead{Cluster}  &\colhead{$N_{\rm{stars}}$} & \colhead{Age}  &\colhead{$\alpha$} &\colhead{$\sigma[\alpha$]} &\colhead{$\delta$} & \colhead{$\sigma[\delta$]} &\colhead{$\mu_{\alpha}$} &\colhead{$\sigma[\mu_{\alpha}$]}&\colhead{$\mu_{\delta}$} &\colhead{$\sigma[\mu_{\delta}$]}&\colhead{$\pi$}  &\colhead{$\sigma[\pi$]}
& \colhead{$N_{\rm{red}}$} &\colhead{Mean} & \colhead{$\sigma$}  & \colhead{$N_{\rm{WD}}$}  \\
 &  & [Myr] & [deg] & [deg] & [deg] & [deg] & [mas/yr] & [mas/yr] & [mas/yr] & [mas/yr] & [mas] & [mas] & & [mag] & [mag] &
}

\startdata
               Alessi 3 & 40 & $600 \pm 100^\dagger$ &  109.117 &  0.664 &  -46.554 &  0.604 &   -9.876 &  0.617 &   11.883 &  0.601 &  3.559 &  0.205 &   31 &  0.098 &  0.021 & 8.8 $\pm{1.7}$    \\
               Alessi 5 &   92 & $80 \pm 20^\dagger$ &  160.802 &  0.885 &  -61.109 &  0.649 &  -15.344 &  0.469 &    2.573 &  0.437 &  2.201 &  1.694 &   55 &  0.256 &  0.014 & 1.6 $\pm{0.7}$    \\
               Alessi 6 &    71 & $30 \pm 10^\dagger$ &   221.96 &  0.253 &   -66.43 &  0.133 &   -7.552 &  0.248 &  -10.301 &  0.288 &  2.326 &  0.345 &   37 &  0.193 &  0.034 & $0 \pm 0.32$    \\
               Alessi 8 &   192 & $100 \pm 30^\dagger$ &  232.624 &  1.402 &  -51.271 &  0.982 &   -5.914 &  0.165 &   -5.689 &  0.163 &  1.479 &  0.152 &   62 &  0.149 &  0.022 &  1.6 $\pm{0.7}$ \\
               Alessi 9 &  - & $281 \pm 135^\dagger$ &  265.908 &  0.217 &  -46.944 &  0.225 &    9.844 &  0.178 &   -8.948 &  0.072 &  4.792 &  0.512 &   29 &   0.07 &  0.019 & $2.3 \pm 0.97$     \\
              Alessi 12 &   182 & $100 \pm 40^\dagger$  &  310.949 &  0.753 &   23.799 &  0.703 &    4.354 &  0.218 &   -4.645 &  0.184 &  1.848 &  0.445 &  125 &  0.098 &  0.008 & 2.8 $\pm{1.6}$    \\
              Alessi 13 &     - & -&        - &      - &        - &      - &        - &      - &        - &      - &      - &      - &    - &      - &      - &  - \\
              Alessi 19 &    56 & $110 \pm 90^\dagger$ &  275.248 &  2.065 &   11.923 &  1.997 &   -1.015 &  0.158 &   -7.001 &  0.158 &  1.545 &  0.276 &   12 &  0.115 &  0.059 & 0.54 $\pm{0.68}$    \\
\enddata
{\medskip {\bf Notes.} $\dagger$ denotes an age from \cite{2020arXiv200407274C}, $\ddagger$ indicates manual age determination from this work, and * indicates that no age determination was possible using our methods. In cases where we could not effectively determine an age estimate expected $N_{\rm{WD}}$ is similarly undetermined. Convert E(Bp-Rp) to E(B-V) using E(B-V) = $ 0.771$E(Bp-Rp) or to G-band extinction using $A_{\rm G} = 2.059$E(Bp-Rp) \citep{2012MNRAS.427..127B}, but see also \citep{2018MNRAS.479L.102C}.
}

\end{deluxetable*}
\end{longrotatetable}

\begin{deluxetable*}{lccccccc}
\tablecaption{WDs appearing in \cite{2018ApJ...866...21C} but not meeting narrow search criteria\label{tableWDs-excluded}}
\tablewidth{700pt}
\tabletypesize{\scriptsize}
\tablehead{
\\
\colhead{Name} &
\colhead{Gaia Source ID} &
\colhead{Parallax} &
\colhead{Pmra} &
\colhead{Pmdec} &
\colhead{G$_{obs}$} &
\colhead{(Bp - Rp)$_{obs}$} &
\colhead{Comments}
\\
 &
 &
[mas] &
[mas/yr] &
[mas/yr] &
[mag] &
 &
\\ 
}
\startdata
GD50	  & 3251244858154433536	& 32.037$\pm$0.060	& 84.429$\pm$0.110	& -162.959$\pm$0.088	& 14.035$\pm$0.001	& -0.555$\pm$0.007 & Not in search [1] \\
Hyeggr29  &	45980377978968064	& 19.940$\pm$0.093 & 	111.454$\pm$0.268 & 	-22.021$\pm$0.146 & 	15.378$\pm$0.001 & 	-0.158$\pm$0.005 & Not identified [2] \\
Hyeggr316 &	3308403897837092992	& 21.732$\pm$0.056 & 	95.276$\pm$0.110 & 	-20.687$\pm$0.051 & 	14.932$\pm$0.001 & 	-0.196$\pm$0.005 & Not identified [2]  \\
Hygd52	  & 218783542413339648	& 23.558$\pm$0.046 & 	145.993$\pm$0.098 & 	-77.695$\pm$0.067 &	15.202$\pm$0.001 & 	-0.104$\pm$0.003 & Not identified [2] \\ 
Hygd74	& 958262527212954752	& 17.189$\pm$0.063	 & -10.587$\pm$0.085 & 	-114.047$\pm$0.078 & 	15.000$\pm$0.001 & 	-0.226$\pm$0.008 &  Not identified [2] \\
Hygd77	& 966561537900256512	& 24.857$\pm$0.044 & 	-16.619$\pm$0.072	 & -170.643$\pm$0.064 & 	14.845$\pm$0.001 & 	-0.100$\pm$0.006 & Not identified [2]\\
Hyhg7-85	& 39305036729495936	& 24.053$\pm$0.054 & 	141.190$\pm$0.107	 & -24.070$\pm$0.084 & 	15.039$\pm$0.001 & 	-0.128$\pm$0.003 & Not identified [2] \\
Hyhz14	& 3294248609046258048	& 20.247$\pm$0.051 & 	91.354$\pm$0.094 & 	-10.643$\pm$0.050 & 	13.864$\pm$0.001 & 	-0.439$\pm$0.007 & Not identified [2] \\
\enddata
{\medskip {\bf Notes.} [1] GD50 is not expected to be part of any cluster included in the narrow search. [2] The Hyades was not identified in proper motion space using our methodology as it has such a large angular extent on the sky. The 11 WDs in this cluster were still included in the IFMR as extensive data has established cluster membership, e.g. see \cite{2018MNRAS.480.3170S}. [3] Do not pass cuts in proper motion and parallax, but with such large errors in the astrometry, we cannot preclude membership. [4] Negative parallax leads to uncertain cluster member candidacy. [5] Appear to be evolving off the main sequence and do not look to be WDs at all.}
\end{deluxetable*}

\end{document}